\LoadClass[12pt]{article}
\RequirePackage{cite}
\RequirePackage{times}
\RequirePackage{fullpage}
\RequirePackage{ifthen}

\makeatletter  
\renewcommand\@biblabel[1]{#1.}  

\def\@cite#1#2{$^{\mbox{\scriptsize #1\if@tempswa , #2\fi}}$}

\newcommand{\spacing}[1]{\renewcommand{\baselinestretch}{#1}\large\normalsize}
\spacing{1}

\def\@maketitle{%
  \newpage\spacing{1}\setlength{\parskip}{12pt}%
    {\Large\bfseries\noindent\sloppy \textsf{\@title} \par}%
    {\noindent\sloppy \@author}%
}

\newenvironment{affiliations}{%
    \setcounter{enumi}{1}%
    \setlength{\parindent}{0in}%
    \slshape\sloppy%
    \begin{list}{\upshape$^{\arabic{enumi}}$}{%
        \usecounter{enumi}%
        \setlength{\leftmargin}{0in}%
        \setlength{\topsep}{0in}%
        \setlength{\labelsep}{0in}%
        \setlength{\labelwidth}{0in}%
        \setlength{\listparindent}{0in}%
        \setlength{\itemsep}{0ex}%
        \setlength{\parsep}{0in}%
        }
    }{\end{list}\par\vspace{12pt}}

\renewenvironment{abstract}{%
    \setlength{\parindent}{0in}%
    \setlength{\parskip}{0in}%
    \bfseries%
    }{\par\vspace{-6pt}}

\newenvironment{methods}{%
    \section*{Methods}%
    \setlength{\parskip}{12pt}%
    }{}

\newenvironment{addendum}{%
    \setlength{\parindent}{0in}%
    \small%
    \begin{list}{Acknowledgements}{%
        \setlength{\leftmargin}{0in}%
        \setlength{\listparindent}{0in}%
        \setlength{\labelsep}{0em}%
        \setlength{\labelwidth}{0in}%
        \setlength{\itemsep}{12pt}%
        }
    }
    {\end{list}\normalsize}

\newcommand*{\addendumlabel}[1]{\textbf{#1}\hspace{1em}}

\documentclass[12pt]{article}
\usepackage[colorlinks=true,urlcolor=blue,linkcolor=blue,citecolor=blue]{hyperref}
\usepackage{graphicx,caption,subcaption}
\usepackage{multibib}
\usepackage{lscape}
\usepackage{xcolor}
\usepackage{float}
\usepackage{amsmath}
\usepackage{amssymb}
\usepackage{longtable}
\usepackage{graphicx}

\newcommand{\aj}{Astron. J.}

\newcommand{\apj}{Astrophys. J.}
\newcommand{\apjl}{Astrophys. J., Letters}

\title{\bf Combined analysis of the 12.8 and 15$~\mu m$ JWST/MIRI eclipse observations of TRAPPIST-1 b}

\author{Elsa Ducrot $^{1,2}$ \thanks{corresponding author ({\color{blue}elsa.ducrot@cea.fr})} \footnote{These authors contributed equally to this work.},
Pierre-Olivier Lagage $^{2}$ $^{\dagger}$,
Michiel Min $^{3}$,
Michaël Gillon $^{4}$,
Taylor J. Bell $^{5}, {6}$,
Pascal Tremblin $^{7}$,
Thomas Greene $^{6}$,
Achrène Dyrek $^{2}$,
Jeroen Bouwman $^{8}$,
Rens Waters $^{9}, {10}, {3}$,
Manuel G\"udel $^{11}, {12}$,
Thomas Henning $^{8}$,
Bart Vandenbussche $^{13}$, 
Olivier Absil $^{14}$,
David Barrado $^{15}$,
Anthony Boccaletti $^{1}$, 
Alain Coulais $^{2}, {16}$,
Leen Decin $^{13}$,
Billy Edwards $^{3}$,
Ren\'e Gastaud $^{2,17}$,
Alistair Glasse $^{18}$,
Sarah Kendrew $^{19}$,
Goran Olofsson $^{20}$,
Polychronis Patapis $^{12}$,
John Pye $^{21}$,
Daniel Rouan $^{1}$,
Niall White\-ford $^{22}$,
Ioannis Argyriou $^{13}$,
Christophe Cossou $^{17}$,
Adrian M. Glauser $^{12}$,
Oliver Krause $^{8}$,
Fred Lahuis $^{23}$,
Pierre Royer $^{13}$,
Silvia Scheithauer $^{8}$,
Luis Colina $^{15}$,
Ewine F.\ van Dishoeck $^{24}$,
G\"oran Ostlin $^{25}$,
Tom P.\ Ray $^{26}$,
Gillian Wright $^{18}$
}

\renewcommand{\figurename}{\hspace{-4pt}}
\renewcommand{\thefigure}{Fig.~\arabic{figure}}
\renewcommand{\theHfigure}{Fig.~\arabic{figure}}
\renewcommand{\tablename}{\hspace{-4pt}}
\renewcommand{\thetable}{Table \arabic{table}}
\renewcommand{\theHtable}{Table \arabic{table}}
\begin{document} 

\maketitle

\begin{affiliations}
        {
          \item  LESIA, Observatoire de Paris, CNRS, Universit\'e Paris Diderot, Universit\'e Pierre et Marie Curie, 5 place Jules Janssen, 92190 Meudon, France. \label{LESIA}
          \item Universit\'e Paris-Saclay, Universit\'e Paris\textbf{} Cit\'e, CEA, CNRS, AIM, F-91191 Gif-sur-Yvette, France \label{cea}
         \item SRON Netherlands Institute for Space Research, Niels Bohrweg 4, 2333 CA Leiden, the Netherlands. \label{SRON}
        \item Astrobiology Research Unit, University of Liège, Allée du 6 août 19, 4000 Liège, Belgium \label{uliege}
        \item Bay Area Environmental Research Institute, NASA's Ames Research Center, M.S. 245-6, Moffett Field, 94035, CA, USA. \label{bay_area}
        \item Space Science and Astrobiology Division, NASA’s Ames Research Center, M.S. 245-6, Moffett Field, 94035, CA, USA. \label{Ames}
        \item Universit\'e Paris-Saclay, UVSQ, CNRS, CEA, Maison de la Simulation, 91191, Gif-sur-Yvette, France.\label{cea_simu}
        \item Max-Planck-Institut f\"ur Astronomie (MPIA), K\"onigstuhl 17, 69117 Heidelberg, Germany. \label{max_planck}
        \item Department of Astrophysics/IMAPP, Radboud University, PO Box 9010, 6500 GL Nijmegen, the Netherlands.\label{uRadboud}
        \item HFML - FELIX. Radboud University PO box 9010, 6500 GL Nijmegen, the Netherlands. \label{uRadboud2}
         \item Department of Astrophysics, University of Vienna, T\"urkenschanzstrasse 17, 1180 Vienna, Austria.\label{uvienna}
        \item ETH Z\"urich, Institute for Particle Physics and Astrophysics, Wolfgang-Pauli-Strasse 27, 8093 Z\"urich, Switzerland.\label{ETH}
        \item Institute of Astronomy, KU Leuven, Celestijnenlaan 200D, 3001 Leuven, Belgium. \label{uLeuven}
        \item STAR Institute, Universit\'e de Li\`ege, All\'ee du Six Ao\^ut 19c, 4000 Li\`ege, Belgium.  \label{uLiege2}
        \item Centro de Astrobiología (CAB), CSIC-INTA, ESAC Campus, Camino Bajo del Castillo s/n, 28692 Villanueva de la Ca\~nada, Madrid, Spain.\label{INTA}
        \item  LERMA, Observatoire de Paris, Universit\'e PSL, Sorbonne Universit\'e, CNRS, Paris, France. \label{LERMA}
        \item  Universit\'e Paris-Saclay, CEA, IRFU, 91191, Gif-sur-Yvette, France\label{DEDIP}
        \item  UK Astronomy Technology Centre, Royal Observatory, Blackford Hill, Edinburgh EH9 3HJ, UK. \label{ROE}
        \item  European Space Agency, Space Telescope Science Institute, Baltimore, Maryland, USA.  \label{STScI}
         \item Department of Astronomy, Stockholm University, AlbaNova University Center, 10691 Stockholm, Sweden.  \label{uStockholm}
        \item  School of Physics \& Astronomy, Space Park Leicester, University of Leicester, 92 Corporation Road, Leicester, LE4 5SP, UK.  \label{uLeicester}
        \item Department of Astrophysics, American Museum of Natural History, New York, NY 10024, USA.\label{AMNH}
        \item SRON Netherlands Institute for Space Research, PO Box 800, 9700 AV, Groningen, The Netherlands. \label{SRON_Groningen}
        \item Leiden Observatory, Leiden University, P.O. Box 9513, 2300 RA Leiden, the Netherlands. \label{uLeiden2}
        \item Department of Astronomy, Oskar Klein Centre, Stockholm University, 106 91 Stockholm, Sweden. \label{uStockholm2}
        \item School of Cosmic Physics, Dublin Institute for Advanced Studies, 31 Fitzwilliam Place, Dublin, D02 XF86, Ireland. \label{Dublin}
        }

\end{affiliations}

\maketitle

\setlength{\parskip}{10pt}

\bigskip


\begin{abstract}
{\bf 

The first JWST/MIRI photometric observations of TRAPPIST-1 b allowed for the detection of the thermal emission of the planet at 15 $\mu m$, suggesting that the planet could be a bare rock with a zero albedo and no redistribution of heat. These observations at 15 $\mu m$ were acquired as part of GTO time that included a twin program at 12.8 $\mu m$ in order to have a measurement in and outside the CO$_2$ absorption band. Here we present five new occultations of TRAPPIST-1 b observed with MIRI in an additional photometric band at 12.8 $\mu m$.  We perform a global fit of the 10 eclipses and derive a planet-to-star flux ratio and 1-$\sigma$ error of 452 $\pm$ 86 ppm and 775 $\pm$ 90 ppm at 12.8 $\mu m$ and 15 $\mu m$, respectively. 

We find that two main scenarios emerge. An airless planet model with an unweathered (fresh) ultramafic surface, that could be indicative of relatively recent geological processes fits well the data. Alternatively, a thick, pure-CO2 atmosphere with photochemical hazes that create a temperature inversion and result in the CO2 feature being seen in emission also works, although with some caveats.
Our results highlight the challenges in accurately determining a planet's atmospheric or surface nature solely from broadband filter measurements of its emission, but also point towards two very interesting scenarios that will be further investigated with the forthcoming phase curve of TRAPPIST-1 b.

}
\end{abstract}

\noindent
\section{Main}

\subsection{Introduction}
\noindent 
At the time of writing, no atmosphere has been detected around any known temperate (0.1$S_{Earth}$ $<$ $S_p$ $<$ 5$S_{Earth}$) rocky planet outside of our solar system. However, with the James Webb Space Telescope (JWST) it is now possible to search for atmospheres around such planets, notably around the latest red dwarfs (spectral types later than M6, and effective temperature less than 3000 K \cite{kirkpatrick_coolest_1997}), also called ultra-cool dwarf stars. The small sizes of such ultra-cool dwarfs yield favorable planet-to-star radius ratios, enabling atmospheric studies of terrestrial planets. Among the $\simeq$5500 exoplanets that have been confirmed so far, only 17 of them can be classified as likely rocky, and orbiting ultra-cool dwarf stars \cite{anglada-escude_terrestrial_2016,zechmeister_carmenes_2019,gillon_seven_2017, ment_second_2019, dreizler_reddots_2020,peterson_temperate_2023,vanderspek_tess_2019}. In particular, 7 of these 17 planets belong to the same system, the TRAPPIST-1 system. Due to their transiting nature, combined with the infrared brightness and the small size of their host star, as well as the resonant architecture of the system that allows mass measurements via the transit timing variations (TTV) method, the TRAPPIST-1 planets are currently the best known terrestrial planets beyond our own solar system \cite{agol_refining_2021}. For all these reasons, TRAPPIST-1 has been identified as a prime target for JWST \cite{gillon_trappist-1_2020,morley_observing_2017}, as emphasized by the ten programs dedicated to the characterization of the system in the first three years of JWST's operation. These programs are the following: GTO 1177 (PI: Greene), GTO 1279 (PI: Lagage), GTO 1201 (PI: Lafranière), GTO 1331 (PI: Lewis), GO 2589 (PI: Lim), GO 2304 (PI: Kreidberg), GO 2420 (PI: Rathcke), GO 1981 (PIs: Lustig-Yaeger \& Stevenson), GO 3077 (PIs: Gillon \& Ducrot), GO 5191 (PIs: Ducrot \& Lagage). This represents a total of $\simeq$290 hours of observation. Among these ten programs, three were dedicated to the observation of the secondary eclipses (when the exoplanet moves out of sight behind the star) of the two innermost planets, TRAPPIST-1 b and c. As part of the Guaranteed Time Observation (GTO) program \#1177 and Guest Observer (GO) program \#2304, five eclipses of TRAPPIST-1 b and four eclipses of TRAPPIST-1 c were observed using the MIRI imaging F1500W filter. For planet b, the brightness temperature at 15 $\mu m$ derived from the secondary eclipses suggested that the planet has little to no planetary atmosphere redistributing radiation from the host star \cite{greene_thermal_2023}. For planet c, the eclipse depth at 15 $\mu m$ suggested that, if the planet had an atmosphere, it was likely thin and not enriched in CO$_2$ \cite{zieba_no_2023}. 

Examining the thermal emission of a planet in the mid-IR as it experiences its secondary eclipse is a powerful way to constrain the presence of an atmosphere \cite{mansfield_identifying_2019,koll_identifying_2019,crossfield_gj_2022}. The depth of the eclipse can be used to  measure the dayside flux of the planet in a specific bandpass. Furthermore, unlike transit spectroscopy, thermal emission remains immune to spectral contamination from the host star (via the Transit Light Source, TLS, effect \cite{rackham_transit_2018}), and the impact of flares and rotational modulation is minimal in the mid-IR and should thus be much less of a concern than for transmission spectroscopy at shorter wavelengths. Cycle 1 observations of TRAPPIST-1 have shown that flare events occur during most transit observations and have intensities up to several thousands of ppm in the near infrared, of similar order as the transit depths signals \cite{howard_characterizing_2023, lim_atmospheric_2023}. These incessant eruptions combined with the TLS effect complicate the interpretation of transmission spectra to the point where significantly more transits than predicted will be needed to detect putative secondary atmospheres on the TRAPPIST-1 planets. For these reasons, and until a systematic method to properly correct for the TLS effect is developed, emission (spectro)-photometry is the best avenue to put constraints on the nature of the TRAPPIST-1 planets.

We present the results from the third TRAPPIST-1 emission program thus far, which was taken as part of the ExO-MIRI consortium (GTO\#1279). This consists of the observation of five eclipses of TRAPPIST-1 b with MIRI using the F1280W filter (spanning wavelengths from 11.588 $\mu m$ to 14.155 $\mu m$), to be compared to the ones obtained with the F1500W filter (spanning wavelengths from 13.527 $\mu m$ to 16.640 $\mu m$) and presented in ref.$~$\cite{greene_thermal_2023}. 

The F1280W filter strikes an appropriate balance between photometric precision and planet-to-star flux contrast, allowing for the detection of the thermal emission of a planet whose expected dayside temperature is $\simeq500$ K, in orbit around an ultra-cool star whose effective temperature is $2566 \pm 26$ K (ref.\cite{agol_refining_2021}). 
The observations took place on November 21 2022, and on July 6, 7, 15, and 23 2023. Due to a change in the observing strategy between the first visit and the four others, the observations had a duration of 3.21 to 3.88 hours, and the eclipses lasted 36 min each; see \ref{tab:observations} for details. The required dates and times of the observations were estimated from the TTV analysis by ref.$~$\cite{agol_refining_2021} and selected so that there was no contamination by simultaneous transits, secondary eclipses or planet-planet occultations\cite{luger_planet-planet_2017} of other planets in the system.

\subsection{Results}

We performed four independent data reductions using \texttt{Eureka!}\cite{bell_eureka_2022} and a custom pipeline; see \ref{tab:reductions}. For each pipeline, light curves were derived using aperture photometry as discussed in the Methods section. The light curves obtained for each visit from the four distinct reduction pipelines lead to consistent photometric precision; see \ref{tab:reductions}.
Three distinct Bayesian data analyses of the light curves were then carried out with various methods for the sampling of the posterior probability distributions of the system's parameters and one more empirical approach was carried out (see Methods section). Each Bayesian analysis included individual fits of each visit, a global fit of the five eclipses, and, for two of them (ED and MG), a generalized global fit including all ten MIRI JWST eclipses of TRAPPIST-1 b (five at 15 $\mu m$ and five at 12.8 $\mu m$). For each Bayesian analysis, we fitted for an eclipse model with a specific parametrisation, as well as for systematic effects which can include a time-dependent polynomial, exponential ramps, and decorrelation against the position and full width half maximum (FWHM) of the point spread function (PSF).

The four distinct reductions lead to consistent light curves for which the root mean square (RMS) of the residuals among all data reductions range between 672 ppm and 963 ppm (10-20\% more precise than reported at 15 $\mu m$ by ref.$~$\cite{zieba_no_2023}).

The eclipse depth of each visit is estimated using distinct approaches with various sampling methods (MCMC or Nested sampling) with different sets of parameters for the astrophysical model and the instrumental systematic model (see \ref{tab:analysis}). 

\begin{figure*}[ht!]
    \centering
    \includegraphics[width=0.99\textwidth]{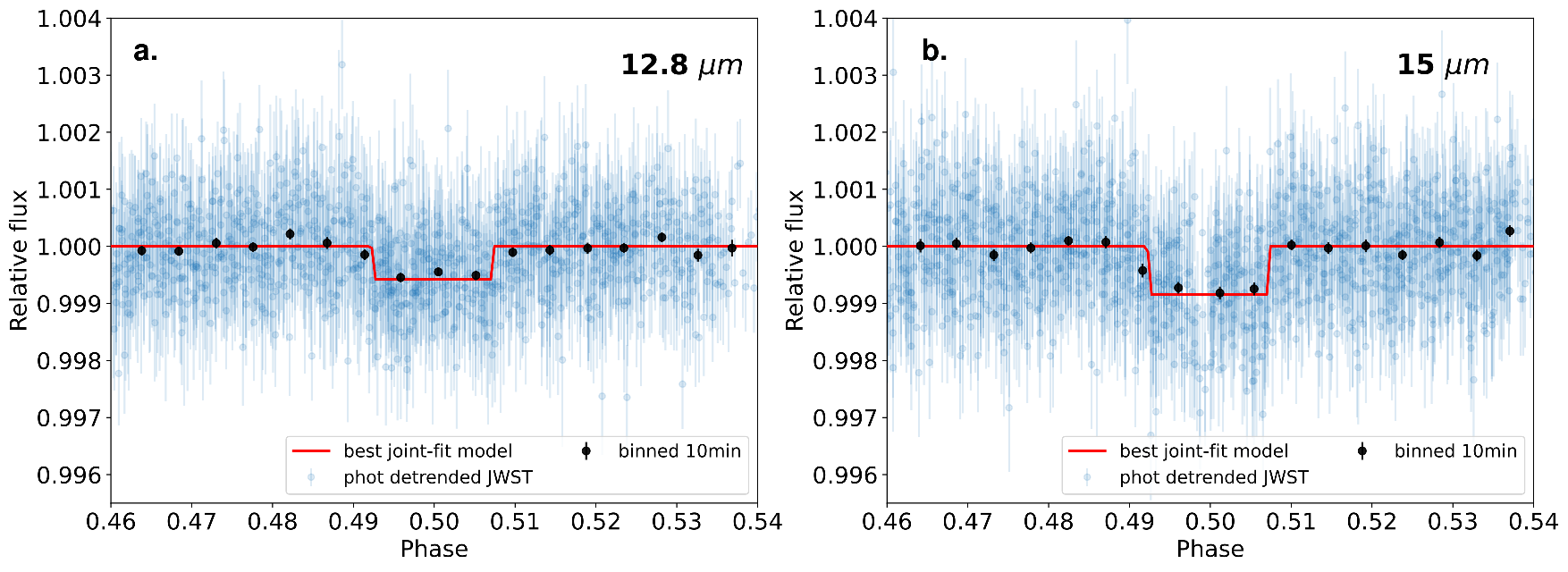}
    \caption{\textbf{Phase folded JWST/MIRI observations of TRAPPIST-1 b.} \textbf{a.} Phase-folded light curve of the secondary eclipse of TRAPPIST-1 b at 12.8 $\mu m$, derived from the observations of 5 eclipses as part of GTO 1279 observations. \textbf{b.} Phase-folded light curve of the secondary eclipse of TRAPPIST-1 b at 15 $\mu m$, derived from re-analyses of the observations of 5 eclipses as part of GTO 1177. These figures are derived from the {\it ``fiducial analysis"} described in the Methods. Blue dots corresponds the corrected flux at the original time sampling, black dots shows the 10min binned corrected flux with an error equal to the standard deviation of the points within the bin, and the red curve show the eclipse model. }
    \label{fig:phase_folded}
\end{figure*}

The derived values from individual fits, joint fits, and generalized fits are presented in \ref{tab:results_depths}. To determine the definitive value of the eclipse depth at 12.8 and 15 $\mu m$,  we performed an additional joint analysis that we call the ``fiducial analysis", for which we used the light curves derived from the reduction with the best photometric precision (MG, see \ref{tab:reductions}) and then apply a different baseline for the systematic model to simplify it, detrending only in time, position and FWHM with a restriction to not use polynomials with degree $>$ 2 to prevent from over-fitting. The detrended phase-folded light curves at 12.8 $\mu m$ and at 15 $\mu m$ from the ``fiducial" global fit of all 10 eclipses are shown in \ref{fig:phase_folded}.
We note that all our analyses are consistent with each other (see Methods). The final eclipse depths, and their 1$\sigma$ uncertainties, are $(F_p/F_{\star})_{b,~12.8~\mu m} = 452 \pm 86~ppm$ and $(F_p/F_{\star})_{b,~15~\mu m} = 775 \pm 90~ppm$. The eclipse depth computed at 15$\mu m$ is in agreement at the 1$\sigma$ level with the one already published ($(F_p/F_{\star})_{b,~15~\mu m,~{\rm Greene+2023}} = 863 \pm 99 ~ppm$\cite{greene_thermal_2023}) as show on \ref{fig:planetary_flux}.

From these eclipse depth values, we derive corresponding blackbody brightness temperatures and 1$\sigma$ uncertainties of $T_{bright,~12.8~\mu m} =424 \pm 28$ K and $T_{bright,~15~\mu m} = 478 \pm 27$ K at 12.8 $\mu m$ and 15 $\mu m$, respectively (see Methods for details on these calculations). Our new estimation of the brightness temperature at 15 $\mu m$, from the joint fit of all 10 eclipses, is thus in agreement at the 1$\sigma$-level with the one derived by ref.$~$\cite{greene_thermal_2023} ($T_{bright,~15~\mu m,~{\rm Greene+2023}} =503 \pm 27$ K).
However, the brightness temperature that we derive at 12.8 $\mu m$ is in disagreement at the 2.1$\sigma$ level with the one expected from the almost null-albedo bare-rock scenario as initially favored by ref.$~$\cite{greene_thermal_2023} (503 $\pm$ 26 K). This result suggests that the nature of TRAPPIST-1 b is potentially more complex than initially expected from the 15 $\mu m$ observations alone. 

\subsection{Discussion}
\begin{figure*}[ht!]
    \centering
    \includegraphics[width=0.7\textwidth]{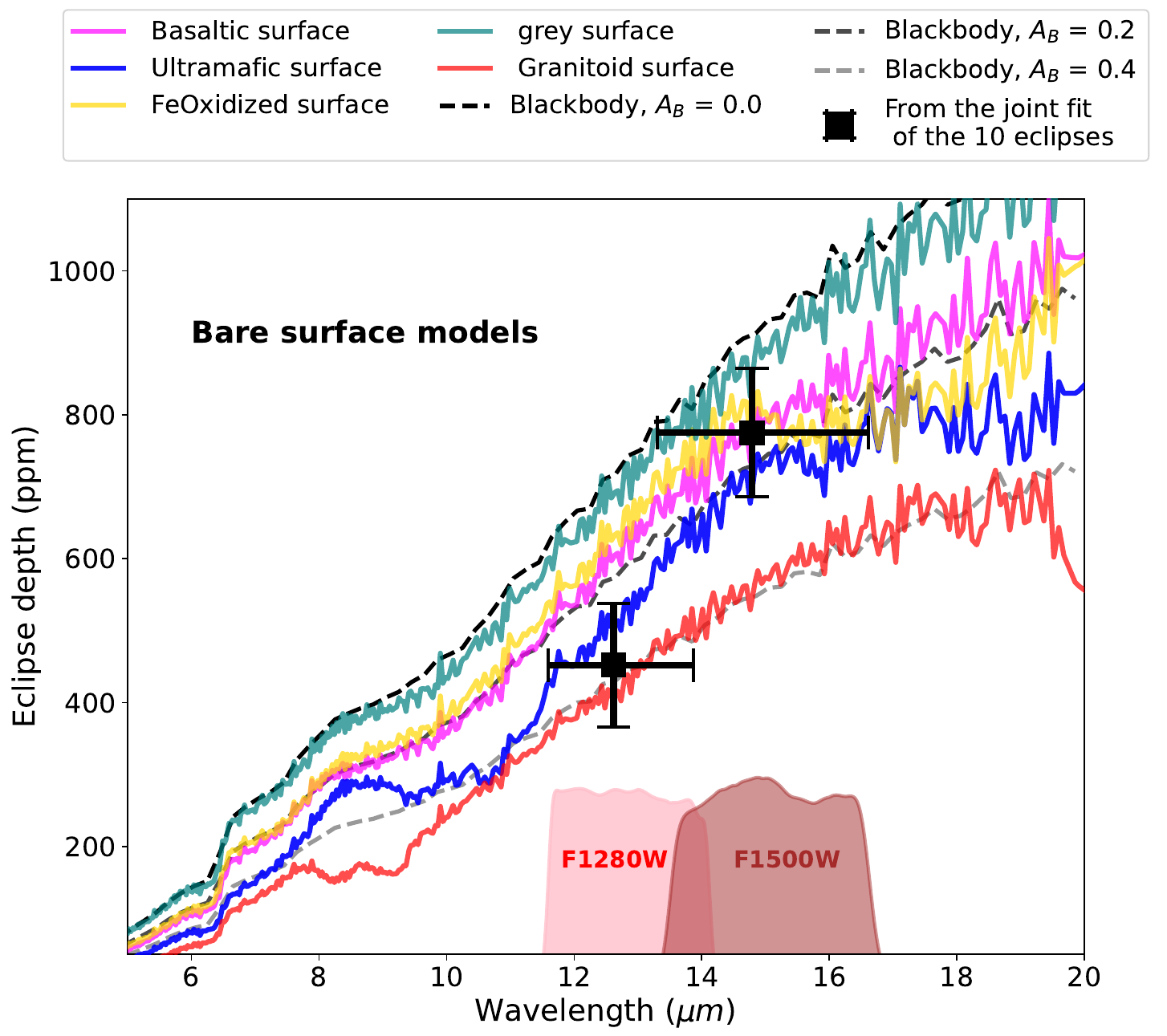}
    \caption{\textbf{TRAPPIST-1 b's emission spectrum compared to bare-surface models.} Measurements of the eclipse depth of TRAPPIST-1 b in the 12.8 $\mu m$ and 15 $\mu m$ bands resulting from 5 visits in each band with their 1$\sigma$ uncertainties from our ``fiducial" joint analysis (MCMC analysis detailed in the Methods section), compared to realistic emission models for bare surface models from ref.~\cite{ih_constraining_2023}. The measurements are centered on the effective wavelength, which is computed by weighting the throughput of the filter with the corrected SPHINX synthetic stellar spectrum (see Methods for details). The error bar in wavelength stands for the width of the filter in each band. 
    Colored markers show the band-integrated depth value for each model. Red and brown filled area show the response of the F1280W and F1500W filters respectively. }
    \label{fig:emission_spectrum}
\end{figure*}

Now that we have measurements of the flux emitted by TRAPPIST-1b in two distinct bands, we can compare them to some surface and atmospheric models. First, we compute the Bond albedo ($A_b$) of the planet by fitting the theoretical emission of a bare rock with no redistribution of heat and albedo as free parameter. We model this theoretical emission spectrum by considering the planetary flux to be a sum of blackbodies calculated for a grid of $T(\theta,\phi)$, where $\theta$ and $\phi$ are the longitude and the latitude, respectively (see Methods for more details). 
The resulting emission spectra for $A_b = 0.,~0.2,~0.4$ are shown in dashed gray lines on \ref{fig:emission_spectrum}. 

The resulting best-fit Bond albedo and its 1$\sigma$ uncertainty is $A_b = 0.19 \pm 0.08$ (see details of how this value is computed in the Methods section). This value is relatively high for a supposedly bare rock that would have endured billions of years of space weathering. Atmosphere-less planets are exposed to the impacts of micrometeorites and stellar irradiation which tend to cover the surface with nanoparticles of metallic iron, resulting in a lowering of the planet's albedo (more details in ref.$~$\cite{lyu_super-earth_2023}). As a comparison, the Bond albedo of Mercury is 0.08 (ref.\cite{madden_catalog_2018, takahashi_mid-infrared_2011}). For close-in planets around active stars, like TRAPPIST-1 b, the timescale of space weathering processes is expected to be much faster, on the order of $10^{2}-10^{3}$ years (ref.\cite{zieba_no_2023}). In addition, ref.~\cite{mansfield_identifying_2019} demonstrated that observations of a bare rock surface lead to the inference of a lower Bond albedo than that of the real surface.

First, we compare our data to the set of plausible surface models (given the level of irradiation received by TRAPPIST-1 b) presented by ref.~\cite{ih_constraining_2023}, including: basaltic, ultramafic, feldspathic, metal-rich, Fe-oxidized (a mixture of basalt and nanophase hematite), and granitoid \cite{hu_theoretical_2012,mansfield_identifying_2019}. We find that the measured planetary flux is most consistent with an ultramafic rock (see Methods) which is composed of 60\% of olivine and 40\% enstatite.

In all the models, the rocks were considered to be geologically fresh. For older surfaces ($\geq 10^{3}$ years) we expect significant impacts of space weathering, resulting in increased eclipse depth in the F1280W and F1500W bands, consequently deteriorating the fit. The presence of fresh ultramafic rock on the surface of TRAPPIST-1 b would be very interesting as it could indicate recent surface processing. Several geological processes such as volcanism resurfacing or crustal reprocessing due to tectonics recycling could explain a more recent surface. TRAPPIST-1 b is expected to experience strong tidal heating due to its close proximity to its star and the perpetual excitation of its eccentricity by the other planets\cite{turbet_modeling_2018}. The planet is also expected to endure induction heating, due to the large magnetic field of its host star\cite{kislyakova_magma_2017}. The combination of these effects should result in a substantial increase of volcanic activity. TRAPPIST-1 b could, therefore, be compared to Io which has a surface so young that it has no impact craters. In that context, the scenario of an airless TRAPPIST-1 b with a young ultramafic surface seems probable. 

Second, we compare our data to plausible atmospheric models.  The initial goal of GTO program \#1177 and GTO program \#1279 was to measure the planetary flux inside and outside the CO$_2$ absorption band (centered around 15 $\mu m$) to investigate the presence of an atmosphere and possibly CO$_2$ at the same time. An atmosphere where the opacity is dominated by greenhouse gases such as CO$_2$ is expected to have a temperature increasing with pressure. Thus, at 15 $\mu$m, where CO$_2$ is strongly absorbing, we would measure the temperature high up in the atmosphere resulting in a lower brightness temperature at 15 $\mu$m than at 12.8 $\mu$m where we can look deeper into the atmosphere. However, our observations suggest the opposite. Within this framework, we explore the possibility that this can be explained by a thermal inversion, where the upper atmosphere is hotter than the atmosphere below, creating a CO$_2$ emission feature instead of a CO$_2$ absorption feature. Such a thermal inversion is observed in the atmosphere of many planets in the Solar System and also in Titan's atmosphere. This inversion is characteristic of a transition to the stratosphere typically around a pressure level of 0.1 bar, and is mainly produced by heating through haze absorption \cite{robinson_common_2014}. The hazes, which are efficient absorbers of stellar radiation, create an inverse greenhouse effect by absorbing the stellar radiation high up in the atmosphere, heating the upper atmosphere and cooling the atmosphere below. Interestingly, the hazes in the atmosphere of Titan are expected to be formed from UV photochemistry. Similarly, hazes could form in the atmosphere of TRAPPIST-1 b through efficient photodissociation resulting from  the much stronger X/Extreme UV (EUV) of TRAPPIST-1 and the proximity of the planet to its star \cite{bourrier_temporal_2017,peacock_predicting_2019,wilson_mega-muscles_2021} (see \ref{fig:irradiance} and Methods).We note that this thermal inversion is not observed in the atmosphere of Venus, which is temperature-wise the most similar to TRAPPIST-1 b in our Solar System. This is likely because of the dominance of highly scattering condensation clouds in the atmosphere of Venus. 

\begin{figure*}[ht!]
    \centering
    \includegraphics[width=\textwidth]{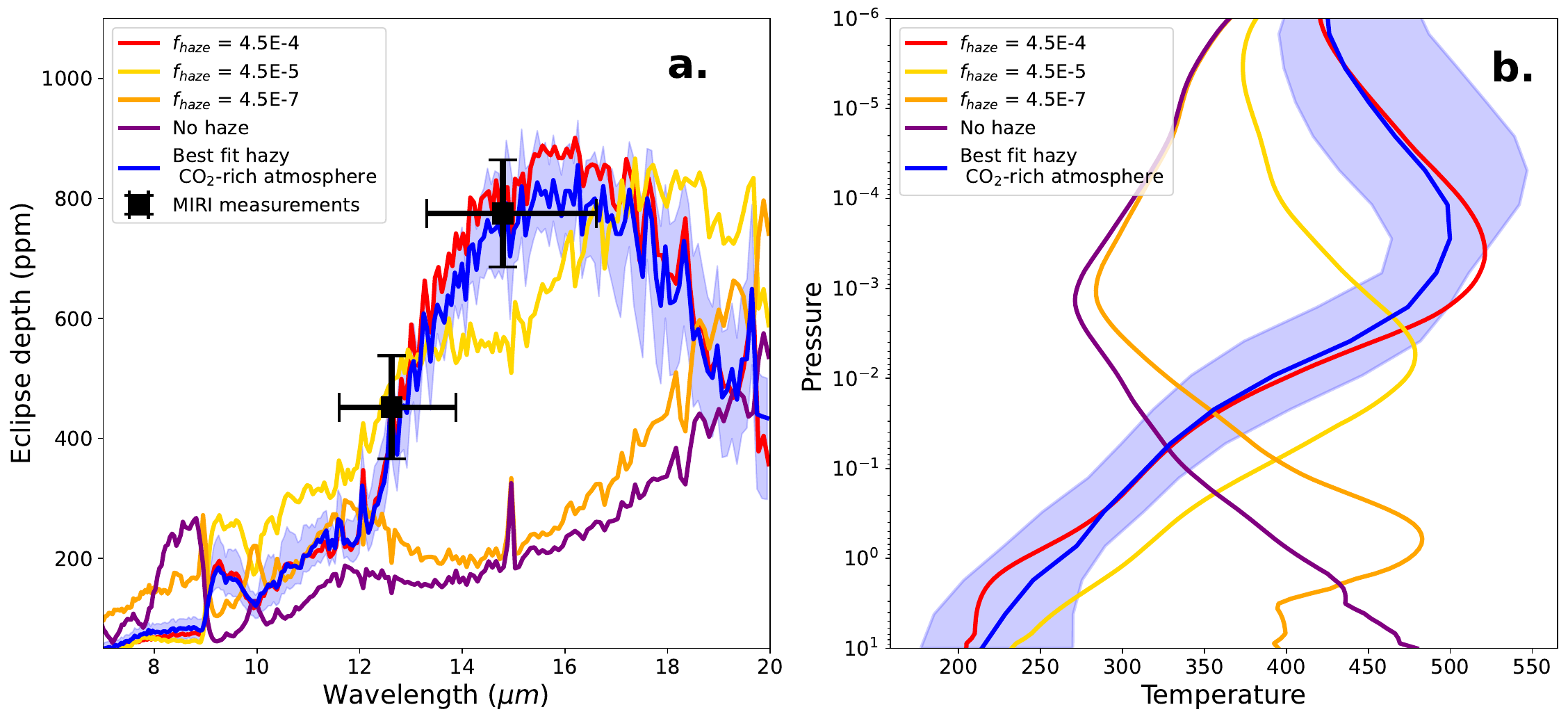}
    \caption{\textbf{Best fit atmosphere model:  an hazy-CO$_2$-rich atmosphere. }Eclipse spectrum (\textbf{a.}) and temperature structure (\textbf{b.}) for the haze atmosphere models. Model calculations are shown for haze mass fractions ($f_{haze}$ = 4.5E-4, $f_{haze}$ =4.5E-5, $f_{haze}$ =4.5E-7) as well as for our best-fit solution (blue line) and its 1$\sigma$ uncertainty interval (blue contour). On panel \textbf{a.}, the MIRI observations, resulting from 5 visits in each bands, are shown by black squares and their 1$\sigma$ errors derived from our ``fiducial" joint analysis are shown by black crosses. The measurements are centered on the effective wavelength, which is computed by weighting the throughput of the filter with the corrected SPHINX synthetic stellar spectrum (see Methods for details). The error bar in wavelength stands for the width of the filter in each band.}
    \label{fig:hazevariation}
\end{figure*}

Using a hazy atmospheric setup inspired by the Titan atmosphere, we show that for a thick, CO$_2$-rich atmosphere with photochemical haze, our model can fit the measurements very well (see \ref{fig:emission_spectrum}). Indeed, the presence of hazes results in a strong thermal inversion causing CO$_2$ to appear in emission (see \ref{fig:hazevariation}). The mixing ratio of hazes seems modest at first glance, but more detailed modeling would be required to assess the feasibility of this level of hazes in the atmosphere. This scenario is interesting as it highlights the importance of thermal inversions in the interpretation of planetary emission spectra. 
However, it comes with some caveats concerning the haze formation and climate stability. For instance, on Titan hydrocarbon hazes are formed through CH$_4$ photodissociation, 
but in the case of a hot, CO$_2$-rich TRAPPIST-1 b the co-existence of CH$_4$ and CO$_2$ is photochemically\cite{thompson_case_2022} and thermodynamically\cite{wogan_abundant_2020} implausible. Yet, ref.~\cite{he_sulfur-driven_2020} showed that the formation of hydrocarbon hazes is experimentally possible without CH$_4$ in a hot, CO$_2$-rich atmosphere that possesses H$_2$S. Although it is unclear yet whether the requisite ~1\% H$_2$S abundances can be maintained by volcanism in comparatively oxidized, CO$_2$-dominated terrestrial atmospheres. 
In that context, this proposed atmospheric scenario is less likely that the bare-surface one, but still worth investigating experimentally and theoretically. 

Our work shows how the measured thermal flux of TRAPPIST-1 b in the F1280W and F1500W MIRI filters can be fitted with either an airless  surface model (potentially with a fresh ultramafic composition) or a thick, CO$_2$-rich atmosphere containing photochemical hazes. Even if the airless scenario is physically more plausible, the existence of an atmosphere around TRAPPIST-1 b still remains undetermined. Importantly, we predict that the simultaneous phase curve measurement at 15 $\mu m$ of TRAPPIST-1 b and c obtained as part of GO program \#3077 will allow us to distinguish between a dense, CO$_2$-rich atmosphere and an airless ultramafic rock (see Methods). 

These results motivate additional observations of TRAPPIST-1 b in emission in distinct bands with JWST, as well as further considerations of thermal inversion in planetary atmospheres in the presence of photochemical hazes. Furthermore, our work highlights how the degeneracy of emission models can complicate the interpretation of broadband measurements. It is currently not possible to reach a definitive conclusion with only two measurements of the planetary thermal flux, a finding which has major implications for other programs seeking to study rocky worlds with emission photometry (e.g., the "hot rocks survey" GO program \#3730, and the "Survey of Rocky Worlds" recently recommended for a DDT Concept\cite{redfield_report_2024}). However, we predict that the combination of broadband emission spectra with photometric phase curve observations offers a robust method to assess the presence of atmospheres around rocky planets.

\pagebreak


\clearpage

\begin{methods}
\renewcommand{\figurename}{\hspace{-4pt}}
\renewcommand{\thefigure}{Extended Data Fig.~\arabic{figure}}
\renewcommand{\theHfigure}{Extended Data Fig.~\arabic{figure}}
\renewcommand{\tablename}{\hspace{-4pt}}
\renewcommand{\thetable}{Extended Data Table \arabic{table}}
\renewcommand{\theHtable}{Extended Data Table \arabic{table}}
\setcounter{figure}{0}
\setcounter{table}{0}

\section*{Observations}\label{Methods:sec:observations}
\noindent Observations were performed as part of Guaranteed Time Observations 1279 (PI: Pierre-Olivier Lagage) in November 2022 and July 2023 (see \ref{tab:observations}). The observation strategy used for the first eclipse (in November 2022) was slightly different than the one used for the four others. Initially the strategy was to offset TRAPPIST-1 from the center of the detector to allow for a second star in the field. The justification for this offset was to use this second star for systematic correction. However, as the amplitude of the detector effects observed with MIRI Imaging turned out to be dependent on the flux of the star, this second star was not as useful as expected. Additionally, TRAPPIST-1 was offset to the edge of the detector where the quality of the flat taken at 12.8 $\mu m$ was slightly worse. As this strategy was adding more constraints to the observations for no significant gain, we decided against utilising it for the 4 other eclipses, instead opting to place TRAPPIST-1 at the center of the detector. \ref{fig:strategy} shows the detector's first image for the first and second visit with TRAPPIST-1 indicated in red. 

\section*{Data Reduction}

Four distinct reductions of the data were performed to ensure consistency of the results. A summary of the different settings for each reduction is given in \ref{tab:reductions}, with detailed explanations given in the sections below.

\noindent \textbf{Data Reduction: ED}

\noindent For the first data reduction, we employed the Eureka! pipeline\cite{bell_eureka_2022} from stages 1 to 4. Our process began with the uncal.fits files, utilizing the default jwst pipeline settings. In stage 1, we tried setting the ramp-fitting weighting parameters to default and then uniform to see if it improved the root mean square (rms) of our residuals. The difference being marginal, we decided to keep the default ramp fitting weighting setup, assigning more weight to the fist and last groups.
In stage 2, we deactivated the photom step. Moving on to stage 3, we defined a subarray region ([642, 742], [460, 560] for the first visit and [642, 742], [460, 560] four the four others), masked pixels flagged in the DQ array, interpolated defective pixels, and conducted aperture photometry on the star. We chose an aperture of 5 pixels for each visit. For each integration, we recorded the center and width of the point spread function (PSF) in the x and y directions by fitting a 2D Gaussian. Background computation involved an annulus with radii of 12--30 pixels (centered on the target), with the result subtracted. We underline that the choice of the background annulus minimally influenced the light curve.
In stage 4, we applied a 4$\sigma$ sigma-clip to identify outliers deviating from the median flux, calculated using a 20-integration-width boxcar filter. 

\noindent \textbf{Data Reduction: MG}

\noindent The second data reduction used the following methodology. The raw uncal.fits files were first calibrated using the first stage of the Eureka! pipeline \cite{bell_eureka_2022} and the second stage of the JWST pipeline. 
A systematic exploration of all the combinations of all Eureka! stage 1 options identified the combination resulting in the
most precise light curves. It corresponded to the default jwst pipeline settings, except for setting the ramp-fitting weighting parameter to uniform, and  the deactivation of the jump correction. 
The rest of the reduction was done using a pipeline coded in \texttt{IRAF} and \texttt{Fortran 2003} which performed the following steps for each image: (1) a change of unit from MJy/sr to recorded electrons, (2) the fit of a 2D Gaussian function on the profile of TRAPPIST-1 to measure the subpixel position of its centroid and its full width at half maximum (FWHM) in both directions, and (3) the measurement of the stellar and background fluxes using circular and annular apertures, respectively,
with \texttt{IRAF/DAOPHOT} \cite{stetson_daophot_1987}. Finally, the resulting light curves were normalized
and outliers were discarded using a 4$\sigma$ clipping with a 20-min moving
median algorithm (average fraction of clipped points: 3.5\%). 
For each visit, the radius of the circular aperture used to measure the stellar flux was optimized by minimizing the standard deviation of the residuals. Apertures of 3.5 pixels were used for all program 1177 data. For program 1279 data, an aperture of 3.5 pixels was used for the first light curve, and of 3.8 pixels for the four others.
The background was measured in an annulus going from 30 to 45 pixels from the PSF center.  
For each stellar flux measurement, the corresponding error was computed taking into account the star and background photon noise, the readout noise, and the dark noise, and assuming a value of 3.1 el/ADU for the gain.

\noindent \textbf{Data Reduction: TJB}

\noindent The third data reduction method started from the \textunderscore uncal.fits files and followed the general procedure of ref.\cite{greene_thermal_2023-1} with a few differences. This reduction used version 0.9 of the \texttt{Eureka!} data analysis pipeline\cite{bell_eureka_2022}, with \texttt{jwst} version 1.10.2 (ref.\cite{bushouse_jwst_2023}), CRDS version 11.17.0 (ref.\cite{crs_developers_crds_2022}), and using the ``jwst\textunderscore 1106.pmap" CRDS context. Stage 1 was run with the firstframe and lastframe steps turned on and the jump rejection threshold set to 10$\sigma$. Stage 2 was run with the photom step turned off. Stage 3 assumed a gain of 3.5 electrons/DN, performed a double-iteration 5$\sigma$ clipping along the entire time axis for each pixel, used a circular annulus with radii of 14 and 34 pixels to remove the background flux, extracted the source flux using a circular aperture with a radius of 8 pixels, and computed the source (x,y) location and PSF-width as a function of time to later use as covariates when fitting the light curve. Finally, in Stage 4 we clipped 3$\sigma$ outliers compared to a 10-integration-wide boxcar-filtered version of the data.

\noindent \textbf{Data Reduction: POL}

\noindent The fourth data reduction and analysis was done in a different manner compared to the three previous ones. The starting point is the uncalibrated data. We have $N_{frames}$x$N_{integrations}$ images with 1032x1024 pixels, where $N_{frames}$ is the number of frames per integration and $N_{integrations}$ is the number of integrations during the exposure. The data are in the form of an array with 4 dimensions: 1032, 1024, $N_{frames}$, $N_{integrations}$; the value of these parameters according to the observations is given in \ref{tab:observations}.   The data reduction is done using a homemade pipeline using the \texttt{IDL} language. The first step is to reduce the size of the images from a 1032x1024 pixels array to a 101x101 pixels array, centered on the maximum of signal of the PSF of the TRAPPIST-1 star. Then we apply the non-linearity correction to each pixel of the $N_{frames}$x$N_{integrations}$ images; for that we use the jwst\_miri\_linearity\_0039.fits file from the JWST Calibration Reference Data System. Because of the so-called last frame effect, we disregard the last frame and consider the signal (in Digital Number) at the Frame number ($N_{frames}$-1); we are working with an array with 3 dimensions: (101,101, $N_{integrations}$).  The jitter amplitude is computer with the \texttt{IDL} \texttt{GCNTRD} routine using a box with a half width of 4 pixels centered on the pixel with the maximum of signal.  As expected, the jitter amplitude is very low (less than 1 mas); there are a number of centroid positions which deviate from the mean position by more than 3$\sigma$, the root cause is not telescope jitter (we do not seen any effect in our data even when there is an high gain-antenna move) but probably cosmic-ray impacts. Indeed, when a cosmic-ray hits the zone of interest, it changes in an inhomogeneous way the signal in the various pixels of the box, changing the position of the centroid. Then we apply an \texttt{aper-ima} \texttt{IDL} function which calculates the sum of the signal in the 57 pixels inside a radius of 4 pixels centered on the pixel with the maximum of signal. This is possible because of the great stability of the telescope. At this point, we have an array with one dimension containing the aperture signal as a function of the integration number.  
To remove outliers, we take advantage of having $N_{frames}$ per integration. We calculate the difference of signal between two consecutive frames, then the mean of the values over time and the noise using the \texttt{meanclip} \texttt{IDL} function. We reject the integrations for which one frame difference was 5$\sigma$ off the mean. We discard 8\% of the frames.

\section*{Data Analysis}

We used four distinct data analysis approaches. For each approach, we performed independent fits of each individual eclipse as well as performing a joint fit of all eclipses at each wavelength. Then, for two of the approaches we performed an additional global joint fit of all 10 eclipses together. The results from these analyses are summarized in \ref{tab:results_depths}.

\noindent \textbf{Data Analysis: ED}

\noindent After acquiring the light curve for each visit in stage 4 of the Eureka! pipeline, we employed the Fortran code \texttt{trafit}, a modified version of the adaptive MCMC code discussed in references \cite{gillon_improved_2012,gillon_search_2014}. This code utilizes the eclipse model from ref.\cite{mandel_analytic_2002} as a photometric time series, multiplied by a baseline model that represents various astrophysical and instrumental systematics contributing to photometric variations. Initially, we individually fitted all visits, exploring a broad range of baseline models to accommodate different sources of flux variation/modulation, such as instrumental and stellar effects. This encompassed polynomials of varying orders concerning time, background, PSF position on the detector (x, y), and PSF width (in x and y).
Upon selecting the baseline, we conducted a preliminary analysis with a single Markov chain comprising 50,000 steps to assess the need for rescaling photometric errors. This involved considering potential underestimation or overestimation of the white noise in each measurement and the presence of time-correlated (red) noise in the light curve. After adjusting the photometric errors, two Markov chains of 100,000 steps each were run to sample the probability density functions of the model parameters and the physical system parameters. The convergence of the MCMC analysis was evaluated using the Gelman and Rubin statistical test.
For each individual analysis, jump parameters with normal distributions were employed: $M_\star$, $R_\star$, $T_{\rm eff}$, [Fe/H], $t_0$, $b$, and $e$. All priors were adopted from refs.\cite{ducrot_trappist-1_2020, agol_refining_2021}. The values of $P$ and $i$ were fixed to the literature values provided in the cited references. The resulting eclipse depths are given in \ref{tab:results_depths}.

We then performed a global analysis with all five visits at 12.8 $\mu m$ and all five visits at 15 $\mu m$. The baseline models, derived from our individual fits for each light curve, were employed. Once again, a preliminary run involving a single chain of 50,000 steps was executed to estimate uncertainty correction factors, which were then applied to the photometric error bars. This was followed by a second run with two chains of 100,000 steps.
The jump parameters mirrored those used in the individual fits, with the exception that $t_0$ was fixed, and allowance was made for Transit Timing Variations (TTVs) to occur for each visit. For each transit TTV, an unconstrained uniform prior was centered on the predicted value from reference\cite{agol_refining_2021}. The Gelman and Rubin statistic was utilized to evaluate the convergence of the fit. \\
Finally, a global analysis with all 10 eclipses was performed were all orbital parameters were common between 12.8 $\mu m$ and 15 $\mu m$, but the depth and the limb darkening parameters were fitted independently. The resulting eclipse depths are given in \ref{tab:results_depths}.

\noindent \textbf{Data Analysis: MG}

\noindent The second data analysis was similar in most respects to the first one, using the Fortran code \texttt{trafit} to analyse individually and globally the light curves. The only differences with the analysis done by ED was that the orbital inclination $i$ was not fixed. It was nevertheless under the constraint of the prior used on the transit impact parameter.  

\noindent \textbf{Data Analysis: TJB}

\noindent For our third data analysis, we continued using the fitting tools in the \texttt{Eureka!} package. For each of our fits, we used a batman eclipse model corrected to account for light-travel time as our astrophysical model. Our astrophysical priors were Normal priors of $P=1.5108794\pm0.000006$ days\cite{agol_refining_2021}, $t_0=2{,}459{,}891.015\pm0.005$ BJD\textunderscore TDB\cite{agol_refining_2021} (Barycentric Julian Date, Barycentric Dynamical Time) which we use to set the time-of-eclipse, and $F_{\rm p}/F_{\rm *}=600\pm2{,}000$ ppm, and we fixed the following parameters which are poorly constrained using eclipse-only observations: $i=89.728^{\circ}$ (ref.\cite{agol_refining_2021}), $a/R_{\rm *}=20.843$ (ref.\cite{agol_refining_2021}), and $e=0$ (not important to fit for since $e$ is very small and we're already fitting for the time-of-eclipse which would be the main impact of a non-zero eccentricity).

Our systematic noise model consisted of a linear trend in time, a linear decorrelation against the PSF width in each of the $x$ and $y$ directions, and an upward exponential ramp and a downward exponential ramp with weakly constrained amplitudes and timescales (since a single exponential ramp was visually insufficient). We also included a Gaussian Process (GP) with a \mbox{Mat\'ern-3/2} kernel (implemented within the celerite package\cite{foreman-mackey_fast_2017}) as a function of time with a uniform prior on the log-amplitude and log-timescale of $\mathcal{U}$(-25,0) and $\mathcal{U}$(-7,0) as otherwise our lightcurve residuals appeared to show significant red noise. Finally, we also included a white-noise error inflation parameter to account for any inaccuracies in the estimated gain as well as the impact of background noise and any other white noise components.

Following the same general procedure as ref.\cite{greene_thermal_2023-1}, we first individually fit each of the new F1280W observations completely independently of each other (fixing the orbital period to $P = 1.510879$ days\cite{agol_refining_2021} to avoid degeneracies between $t_0$ and $P$). Then we performed a fit where we used the same orbital period and time-of-eclipse for all observations, and then finally we performed a fit where we used the same orbital period, time-of-eclipse, and eclipse depth for all observations. For each of our fits, we used the nested sampling algorithm\cite{skilling_nested_2006} from the \texttt{dynesty} package\cite{speagle_dynesty_2020}. For each fit, we used a number of live points equal to $N_{\rm dim}(N_{\rm dim}+1)//2$, where $N_{\rm dim}$ is the number of fitted parameters, and $//$ denotes integer division; we also used `multi' bounds, the `auto' sampling algorithm, and ran the fits until the stopping criterion of $\Delta\log{\hat{\mathcal{Z}}}\leq0.1$ was met. Ultimately, this required about 70 million log-probability evaluations for each of our three runs to reach convergence which took up to 78 hours. Because this method was so computationally intensive, we were not able perform a further fit with both the F1280W and the F1500W eclipses.

\noindent \textbf{Data Analysis: POL}

\noindent The fourth analysis was based on the fact that the time of the eclipse is precisely predicted. Then we just sum the signal in the various integrations occurring during the predicted eclipse duration, narrowed by the uncertainties in the predicted eclipse time ($\pm$ 2 minutes). We take the sum of the signal in the same number of integrations before the eclipse and after the eclipse. The eclipse is rather far from the beginning of the observation and the drift of the detector is smooth, so that we can approximate it by a straight line. Then, from the comparison of the flux before and after the eclipse, we deduce the drop in flux during the eclipse. The difference between this value and the signal during the eclipse measures the depth of the eclipse. To have the value in ppm, we divide by the star flux. To get the star flux we remove the background from the signal. The background represents about 46 percents of the signal at 15 $\mu m$ and about 31 percent at 12.8 $\mu m$. The final value is the mean of the 5 eclipses. 
The noise is measured in the integration windows before, during and after the eclipse and is compared with the expected photons noise (assuming a electronic gain of 3.5 to go from DN to electrons). We then extrapolate the noise as square root of the number of integrations in the windows and then combining quadratically the noises of the 5 eclipses to get the final eclipse depth noise. We also take into account the uncertainty on the star signal, which is negligible compare to the photon noise. Note that when combining the 5 eclipses the noise in the sum of the various eclipses varies as square root of 5, as expected.

\noindent \textbf{Fiducial joint data analysis}

The ``fiducial" joint analysis is the one that lead to the results presented in the main text. This analysis uses the light curve from the MG reduction, as these are the one with the best RMS of the residual before detrending. Then the joint analysis is very similar to the one of ED and MG. The 10 eclipses of TRAPPIST-1 b were fitted with the following jump parameters: $M_\star$, $R_\star$, $T_{\rm eff}$, [Fe/H], $t_0$, $b$, $e$, and TTVs. All priors were adopted from references \cite{ducrot_trappist-1_2020, agol_refining_2021}. The only difference with ED's analysis is in the baseline selection. Baselines include polynomial in time, position and FWHM, with a restriction to not use degree $>$ 2 to prevent from over-fitting.

\noindent \textbf{Stellar flux and brightness temperature calculation}

We conducted measurements of the star's absolute flux density across all calibrated images, utilizing a 25-pixel aperture (similarly to ref.$~$\cite{greene_thermal_2023, zieba_no_2023}). 
Converting these flux densities from MJy$/sr^{-1}$ to $mJy$, we calculated mean values of 3.42 $\pm$ 0.019 $mJy$ and 2.528 $\pm$ 0.022 $mJy$ at 12.8 and 15 $\mu m$, respectively. 
To this error of approximately of 0.7\% and 0.8\%, we added a systematic error of 3\%, reflecting the estimated absolute photometric precision of MIRI as indicated in the most recent report on the temporal behavior of the MIRI Reduced Count Rate. This resulted in a combined error of 0.12 mJy and 0.097 mJy at 12.8 and 15 $\mu m$, respectively. Our measured flux at 15 $\mu m$ is therefore within 1$\sigma$ agreement with the measures of ref.~\cite{greene_thermal_2023} (2.589 $\pm$ 0.078 mJy) and ref.~\cite{zieba_no_2023} (2.559 $\pm$ 0.079 mJy).
We note that to compute the mean stellar flux value at 12.8 $\mu m$, we have used only the visits that were acquired in the same conditions; that is to say visits 2 to 5. As explained previously \ref{Methods:sec:observations}, for the first 12.8 $\mu m$ visit TRAPPIST-1 was positioned in the bottom-left corner of the full frame and not in the center like for all the other observations (including the ones from GTO 1177 and GO 2304), see \ref{fig:strategy}. This strategy was supposed to allow for another star in the field for potential comparisons, however the bottom left region is more impacted by the background and flat-fielding corrections. In that regard, we decided not to include the stellar flux measured from the first visit in the calculation of mean for consistency. We also note that we have reduced all the data with the same version of the jwst (CRDS version 7.5.0.3, using the ``jwst\textunderscore1089.pmap" CRDS context) pipeline for all visits (for GTO 1279 and GTO 1177).

Then, by multiplying the measured flux density by the observed occultation depth, we derived a planetary flux density of 1.55 $\pm$ 0.29 $\mu Jy$ at 12.8 $\mu m$ and 2.18 $\pm$ 0.25 $\mu Jy$ at 15 $\mu m$. We show these results compared to the ones from ref.$~$\cite{greene_thermal_2023} in \ref{fig:planetary_flux}. Then, employing Planck's law, we computed the planet's brightness temperature of $T_{bright, 12.8 \mu m}$ = 424 $\pm$ 28 K and $T_{bright, 15 \mu m}$ = 478 $\pm$ 27 K, to be contrasted with an equilibrium temperature of 508 K computed for a planet with zero albedo and zero heat distribution across its surface.

\noindent \textbf{Potential eclipse depth variability} 

\noindent When we analyze each light curve individually, we observe some indicative variability in the eclipse depth at 12.8 $\mu m$ with all data analysis methods (see \ref{tab:results_depths}) but not at 15 $\mu m$ (see \ref{fig:depth_var_dates}). To asses whether this variability is real or not, we performed several statistical tests (Shapiro–Wilk test, Anderson-Darling test, and Monte-Carlo test). Each statistical test reached the same conclusion that this variability is currently not significant, and we can not reject the null hypothesis that all five eclipse depths were drawn from a single Normal distribution based on the current observations only. Hence, additional eclipses at 12.8 $\mu m$ are needed to confirm or reject the depth variability hypothesis. As a thought experiment, if we hypothetically assume that this time-dependent occultation depth variability at 12.8 $\mu m$ for TRAPPIST-1 b is real, this would have very interesting implications. Possible explanations include: 
\vspace{-0.3cm}
\begin{itemize}
    \item The existence of an out-gassing source (volcanoes, geysers etc) that creates gases or dusts that would absorb the planetary thermal flux at 12.8 $\mu m$, but that would then be blown away rapidly by stellar activity (in the time scale of a day as visit 2 and 3 at 12.8 $\mu m$ are only one orbital period apart from each other, see \ref{fig:depth_var_dates}). Interestingly, due to the close proximity to its star and the perpetual excitation of its eccentricity by the other planets, TRAPPIST-1 b is very likely strongly tidally heated \cite{turbet_modeling_2018}, which can lead to strong volcanism.
    \item The existence of photochemical processes between stellar activity and molecules present in a putative atmosphere of TRAPPIST-1 b that results in the creation of molecules or hazes that either absorb or emit strongly at 12.8 $\mu m$ on the timescale of a day.
\end{itemize}

\section*{Modeling of the thermal emission of the planet} 

For the stellar spectrum, we used the SPHINX model grid\cite{iyer_sphinx_2023} interpolated to TRAPPIST-1's parameters\cite{agol_refining_2021, ducrot_trappist-1_2020}. We compared the integrated binned stellar fluxes of this theoretical spectrum weighted by filter F1280W and F1500W and compared them to the measured stellar fluxes (3.42 $\pm$ 0.019 $mJy$ and 2.528 $\pm$ 0.022 $mJy$ at 12.8 and 15 $\mu m$, respectively). Similarly to ref.$~$\cite{ih_constraining_2023}, we find that a $\simeq$7\% correction needs to be applied to the SPHINX synthetic spectrum to match the observations. For the computation of the following emission models (bare surfaces and atmospheres) this corrected SPHINX spectrum was used.

\subsubsection*{Bare surfaces} 

The surface emission models used here are the same as the ones presented in ref.$~$\cite{ih_constraining_2023}. These models were generated using the open source 1D radiative transfer code \texttt{HELIOS}, which computes the temperature-pressure profiles and emission spectra in radiative-convective equilibrium, including the effects of a solid surface\cite{malik_analyzing_2019, whittaker_detectability_2022}. Six different surfaces were modeled: basaltic, ultramafic, feldspathic, metal-rich, Fe-oxidized, and granitoid\cite{hu_theoretical_2012,mansfield_identifying_2019}. The emission spectrum for each scenario was derived from the wavelength-dependent albedo spectrum of the various surfaces. 
For the gray albedo emission model, we do not follow the exact methodology of ref.$~$\cite{ih_constraining_2023}, where a constant temperature is assumed across the entire dayside of the planet. Instead, we modeled the theoretical emission spectrum of an airless planet with a Bond albedo $A_b$ in the absence of heat redistribution by considering the planetary flux $F_p$ to be a sum of black bodies calculated for a grid of $T_{\theta,\phi}$, defined as: 
\begin{equation}
    F_p = \Big(\dfrac{R_p}{d}\Big)^2 \Big(\int_{-\pi/2}^{\pi/2}\int_{-\pi/2}^{\pi/2} \epsilon_\lambda B_\lambda[T(\theta,\phi)] \textrm{cos}\theta \textrm{cos}\phi d\theta d\phi\Big)
\end{equation}
 where $\theta$ and $\phi$ are the longitude and the latitude, respectively, and the emissivity of the planet ($\epsilon_\lambda$) is constant and equal to $(1-A_B)$.

For a wavelength independent albedo, the temperature of the sub-stellar point (at zenith) is independent of that albedo and has a value $T_{\rm dayside,~max}=T_\star\times \sqrt{\frac{R_\star}{a}}$. The temperature then decreases with increasing latitude and longitude up to the terminator and is fixed to $45$ K on the nightside (see left panel of \ref{fig:besfit_albedo}). From this theoretical definition, we perform a Bayesian analysis, varying the radius of the planet and its albedo, simultaneously fitting the transit spectrum and the eclipse depth. As constraints on the model we use the MIRI eclipse measurements detailed above and the NIRISS/SOSS transit transmission spectrum as presented in ref.$~$\cite{lim_jwst_2023}. From this, we derive the value of the Bond albedo that best fits the measurements of the eclipse depths at 12.8 and 15 $\mu m$ (see right panel of \ref{fig:besfit_albedo}). The resulting best-fit of the NIRISS transmission spectrum is shown on \ref{fig:fit_transmission}).

We derive $A_B = 0.19 \pm 0.08$, which is a relatively higher value than expected for an atmosphere-less planet exposed very frequently to important stellar activity, like TRAPPIST-1 b. Space weathering should drastically lower the albedo of the planet and reduce the spectral contrast via the creation of nanometer-scale iron particles\cite{domingue_mercurys_2014}. We note that, considering the error that we derive on the albedo, we do not have evidence against a grey albedo and have mild evidence (1.8 sigma) against a null-albedo black surface. 

Additionally, we compared our measurement to the recently-published bare-surface models of ref.~\cite{ih_constraining_2023}. For this comparison, we did not fit any parameters and only compared the predictions to the observations and computed the reduced-$\chi^2$ of each surface model for 2 data points and 1 fitted parameter (see \ref{tab:chi2_surfaces}). We conclude that the ultramafic surface model is favoured.
Alternatively, a highly weathered granitoid surface could be also be a possibility but this is out of the scope of this paper as we did not modeled the effect of space weathering on the bare-surfaces emission spectra.

\subsubsection*{Atmospheric models} 

\noindent \textbf{1D atmosphere with full heat redistribution}

For the atmosphere model of TRAPPIST-1b, we initially use the setup from ref.$~$\cite{malik_analyzing_2019} with a pure CO$_2$ atmosphere with full heat redistribution. We take the surface pressure to be 10\,bar and a gray surface albedo of 0.1. The only change with respect to the setup from ref.$~$\cite{malik_analyzing_2019} is that we mix in hydrocarbon haze particles. This is a rather extreme setup in terms of heat redistribution efficiency, but we use it as a proof of concept to show the dependence of the atmospheric flux with haze content. For the optical properties of the hydrocarbon haze particles, we use the optEC$_{(s)}$ model as derived by ref.$~$\cite{jones_advanced_2013} together with Mie theory to convert the refractive index into absorption and scattering cross sections. In this model, the properties of the hydrocarbon particles are modeled as a function of particle size and band gap, $E_g$. We fix the size of the particles to be 50\,nm, consistent with findings for haze production from CO$_2$ in the laboratory \cite{he_laboratory_2018}.

We compute the temperature structure of the atmosphere using the radiative transfer code \texttt{ARCiS} \cite{min_arcis_2020, chubb_exoplanet_2022}. Since we consider full heat redistribution in our nominal setup, we do not use the 3D implementation as presented in ref.$~$\cite{chubb_exoplanet_2022} but an isothermal irradiation with a heat redistribution factor of 0.25. As constraints on the model, we use our MIRI eclipse measurements in addition to the NIRISS/SOSS transmission spectrum presented by ref.$~$\cite{lim_jwst_2023}. Our retrieval setup has only three free parameters: the radius $R_p$, the haze mass fraction in the atmosphere $f_{\rm haze}$, and the band gap in the hydrocarbon particles $E_g$. In Fig.~\ref{fig:retrieval_haze} we present the resulting cornerplot, the resulting best fit transmission spectrum compared to the NIRISS measurement is shown on \ref{fig:fit_transmission}. As can be seen, the haze mass fraction is well constrained. Interestingly, the band gap is also constrained. This is likely caused by the variation of the albedo as function of $E_g$. In our simplistic setup, we consider this result to be preliminary, and more detailed haze models in combination with more constraining observations are required to determine the band gap.


Our median probability model has $f_{\rm haze}=4.5\times 10^{-4}$. To show the dependence of the temperature structure in the atmosphere with increasing $f_{\rm haze}$, we computed models with $0$, $10^{-3}$, $10^{-1}$, and $1$ times this median value. The results are shown in Fig.~\ref{fig:hazevariation}. We see that for the model with $f_{\rm haze}=0$ we reproduce the resulting temperature structure as computed by ref.$~$\cite{malik_analyzing_2019}, with the exception of the temperatures close to the surface. The differences close to the surface are likely caused by a different treatment of the surface/atmosphere interaction, i.e. the lower boundary condition in the radiative transfer. In our model we force the surface temperature to be the same as the lowest atmospheric grid cell, while ref.$~$\cite{malik_analyzing_2019} allow the temperature to make a jump at the surface. The most striking result is the rapid formation of a strong temperature inversion when a haze is added to the atmosphere. The haze particles are very good absorbers in the optical but absorb much less efficiently in the infrared. In this way they create an inverse greenhouse effect, cooling the lower part of the atmosphere and capturing the stellar radiation higher up. This thermal inversion creates a CO$_2$ emission feature around 15 $\mu m$, which boosts the flux at these wavelengths at the expense of flux at other wavelengths.

We have performed a comparison between \texttt{ARCiS} and \texttt{ATMO} \cite{tremblin_fingering_2015, drummond_effects_2016} to check the robustness of the temperature inversion produced by the heating from hazes in the upper atmosphere. Both models consider a CO$_2$ atmosphere with a uniform haze layer through the entire atmosphere with a constant albedo of 0.3 and the following parameterization of the haze opacity $\kappa_1/\lambda^2$ with $\kappa_1=0.5$ cm$^2$/g and $\lambda$ the wavelength in unit of $\mu$m. The resulting PT profiles and eclipse depth are shown in Fig.~\ref{fig:comparison}. Both models reproduce a temperature inversion with temperatures of $\sim$ 500~K at pressures of $\sim 10^{-3}$ bar and a deep-atmosphere temperature around $\sim$250~K. The corresponding eclipse depth shows an increase of about a factor $\sim$2 between 12.8 and 15 $\mu m$, in agreement with the observations. Differences remain between the two models, likely because of inconsistent parameters (e.g., for the stellar irradiation). However, the agreement between the two models is good given the uncertainties of the observations and shows that the inversion is produced robustly without the need to fine tune the models. 

\noindent \textbf{3D atmosphere with partial heat redistribution}

The two cases presented above, a bare ultramafic rock without any redistribution of heat and a thick atmosphere with full heat redistribution, both provide a decent representation of the data. In this section we explore the intermediate case. A simple equation that can be used to compute the heat redistribution factor for a given atmospheric surface pressure is given in ref\cite{koll_identifying_2019}. We use that equation to run retrievals where we fix the surface pressure to a value of 0.1, 1 and 10\,bar but we compute the heat redistribution efficiency using the equations from ref.$~$\cite{koll_identifying_2019}. This allows us to construct a 3D model of the atmosphere including the nightside and thereby make a crude approximation of what a phase curve observation of the planet might look like. If the hazes are formed by photochemistry on the dayside it is unknown if they will transfer to the nightside. Here we assume they do, so we keep $f_{\rm haze}$ constant. This will cause the thermal inversion to also transport to the nightside. Even though there are many uncertainties in this computation, we consider it a useful exercise to study the potential differences in observed phase curve. 

We start our procedure from the median probability model from the previous section. The value for $E_g$ remains fixed while the values for $R_p$, $f_{\rm haze}$ and the surface albedo are varied to optimize the model with respect to the JWST transit and eclipse data.
We construct the 3D atmospheric model using the diffusion concept \cite{chubb_exoplanet_2022} and we compute the strength of the heat diffusion to the nightside by enforcing the night-to-day ratio computed from the equations from ref.$~$\cite{koll_identifying_2019}.\\
The equations for heat redistribution depend on the long wavelength optical depth. Therefore, we iterate on that to find the heat redistribution efficiency. For an atmosphere of 0.1, 1 and 10 bar we find a heat redistribution factor of 0.62, 0.44, and 0.28 respectively, where lower values mean more efficient heat transport. In ref.$~$\cite{chubb_exoplanet_2022} the heat redistribution is defined in terms of the integrated flux on the nightside divided by the integrated flux on the dayside. This parameter, the night-to-day ratio, can also be computed following the equations from ref.$~$\cite{koll_identifying_2019}. It has a value of 0.05, 0.37, and 0.85 for the 0.1, 1 and 10 bar model respectively. For this definition of the heat redistribution higher values correspond to more efficient transport.

In Fig.~\ref{fig:phasecurves_haze} we show the resulting phase curves of TRAPPIST-1 b in the $15\,\mu$m filter for the different atmospheric pressures. We also show the binned simulated MIRI F1500W data that we constructed based on the performances obtained from the reduction of existing 15$\mu m$ observations of TRAPPIST-1 taken as part of GTO program \#1177. As can be seen we should be able to constrain the heat redistribution efficiency, and thereby the surface pressure, by observing the infrared phase curve. Note that if there are no photochemical hazes on the nightside of the planet, it is not given that the nightside has a thermal inversion like the dayside. Therefore, it might be that even with a thick atmosphere, the nightside flux is lower than predicted by our simple model.

\noindent \textbf{Comparison to solar system bodies}

The photochemical haze model presented above is inspired by the hazes in the atmosphere of Titan. Photochemical hazes are predicted to be important in atmospheres of moderately warm exoplanets as well\cite{kawashima_theoretical_2018}. 
TRAPPIST-1 b orbits a star very different from the Sun (see Fig.~\ref{fig:irradiance}). This makes direct comparison to Solar System planets difficult. In terms of total irradiation, TRAPPIST-1 b lies between Mercury, which receives about twice as much radiation, and Venus, which receives about half that of TRAPPIST-1 b. However, because the central star is so different from the Sun the UV flux received by TRAPPIST-1 b does not have any analogues in the solar system. If we integrate the stellar flux between 100 and 400\,nm, using the SED from the Mega MUSCLES project\cite{wilson_mega-muscles_2021}, we find that TRAPPIST-1 b receives about 93\% of the UV radiation that Titan receives from the Sun (see red patch on \ref{fig:irradiance}).

However, if we distinguish the EUV and near-UV contribution of the full UV spectrum, the comparison does not stand anymore. In particular, from a EUV point of view, the TRAPPIST-1 planets receive large EUV irradiations (at least $10^{3}$ x Titan’s flux\cite{turbet_modeling_2018}). Such strong EUV can alter the atmospheres though atmospheric escape\cite{luger_extreme_2015} and can also intensify photochemical reactions in the upper atmosphere resulting in the formation and accumulation of high-altitude organic haze\cite{turbet_modeling_2018}. In particular, the photodissociation rate of CH$_4$ (maximized in the 20-150 nm wavelength range) is expected to be higher\cite{hu_o2-_2020}. We also note that the photodissociation of CO$_2$ happens in the 120–180 nm spectral region. However, the photolysis of CO$_2$ creates oxygen radicals that can lead to the destruction of hazes via the oxidization of hydrocarbon photochemical products\cite{arney_pale_2017}. The currently most updated UV spectrum of TRAPPIST-1\cite{wilson_mega-muscles_2021} shows a drop of flux between 100 and 300 nm which could suggest that the photodissociation of CO$_2$, and therefore the creation oxygen radicals, is not optimal on TRAPPIST-1 planets, such that the presence of hazes might remains possible even in the presence of CO$_2$.

Finally, concerning the thermal inversion resulting from the presence of hazes, the feedback of the absorption properties on the atmosphere is discussed in detail in ref.$~$\cite{robinson_common_2014}. In that paper, a model is presented explaining the thermal inversion observed in e.g. Titan. We note that in our model the thermal inversion does extend to higher pressures because we have a homogenous mixing efficiency of the hazes throughout the entire atmosphere opposed to a more realistic layer of haze. When more observations are available more realistic haze models should be explored to investigate the haze properties and hopefully compare them to the hazes found in the atmospheres of solar system bodies e.g. Titan and Venus.



\clearpage
\section*{Data Availability}
The data used in this paper are associated with JWST GTO programs 1177 (PI: T. Greene) and GTO 1279 (PI: P-O. Lagage) and are publicly available from the Mikulski Archive for Space Telescopes (\url{https://mast.stsci.edu}) on \href{https://mast.stsci.edu/search/ui/#/jwst/results?resolve=true\&target=TRAPPIST-1\&program\_id=1177,\%201279\&cycles=1\&radius=3\&radius\_units=arcminutes&useStore=false&search\_key=c0d442ccd60ea}{this link}. Additional source data, tables and figures from this work are archived on Zenodo at \url{https://zenodo.org/records/13385020}.

\section*{Code Availability}

This work was performed using the following codes to process, extract, reduce and analyse the data: STScI's JWST Calibration pipeline\cite{bushouse_jwst_2023}, \texttt{Eureka!}\cite{bell_eureka_2022}, \texttt{trafit}\cite{gillon_improved_2012} \texttt{starry}\cite{luger_starry_2019}, \texttt{exoplanet}\cite{foreman-mackey_exoplanet_2021}, \texttt{PyMC3}\cite{salvatier_probabilistic_2016}, \texttt{emcee}\cite{foreman-mackey_emcee_2013}, \texttt{dynesty}\cite{speagle_dynesty_2020}, \texttt{numpy}\cite{harris_array_2020}, \texttt{astropy}\cite{robitaille_astropy_2013}, and \texttt{matplotlib}\cite{hunter_matplotlib_2007}.

\begin{addendum}

 \item[Acknowledgments]

We thank Martin Turbet, and Franck Selsis for very useful discussion regarding the modeling of a putative atmosphere of TRAPPIST-1 b and its likelihood. We thank Benjamin Charnay for fruitful discussion regarding photochemical hazes formation processes and comparison with Titan. We thank Jegug Ih for sharing their bare surface models published in ref.\cite{ih_constraining_2023} and for useful discussion regarding the calculation of the Bond albedo. We thank Olivia Lim for sharing their transmission spectra of TRAPPIST-1 b obtained from the reduction and analysis of JWST/NIRISS data and published in their paper ref.\cite{lim_atmospheric_2023}. We thank Eric Agol for his help on planetary flux estimation equations. Finally, we thank Aishwarya Iyer for providing with the sphinx stellar spectrum model extrapolated on TRAPPIST-1's parameters. \\
This work is based on observations made with the NASA/ESA/CSA JWST. The data were obtained from the Mikulski Archive for Space Telescopes at the Space Telescope Science Institute, which is operated by the Association of Universities for Research in Astronomy, Inc., under NASA contract NAS 5-03127 for JWST. These observations are associated with program 1279. MIRI draws on the scientific and technical expertise of the following organisations: NASA Ames Research Center, USA; Airbus Defence and Space, UK; CEA-Irfu, Saclay, France; Centre Spatial de Li\`ege, Belgium; Consejo Superior de Investigaciones Cient\'ificas, Spain; Carl Zeiss Optronics, Germany; Chalmers University of Technology, Sweden; Danish Space Research Institute, Denmark; Dublin Institute for Advanced Studies, Ireland; European Space Agency, Netherlands; ETCA, Belgium; ETH Zurich, Switzerland; Goddard Space Flight Center, USA; Institut d’Astrophysique Spatiale, France; Instituto Nacional de T\'ecnica Aeroespacial, Spain; Institute for Astronomy, Edinburgh, UK; Jet Propulsion Laboratory, USA; Laboratoire d’Astrophysique de Marseille (LAM), France; Leiden University, Netherlands; NOVA Opt-IR group at Dwingeloo, Netherlands; Northrop Grumman, USA; Max-Planck-Institut für Astronomie (MPIA), Heidelberg, Germany; Laboratoire d’Etudes Spatiales et d’Instrumentation en Astrophysique (LESIA), France; Paul Scherrer Institut, Switzerland; Raytheon Vision Systems, USA; RUAG Aerospace, Switzerland; Rutherford Appleton Laboratory (RAL Space), UK; Space Telescope Science Institute, USA; Toegepast Natuurwetenschappelijk Onderzoek (TNO-TPD), Netherlands; UK Astronomy Technology Centre, UK; University College London, UK; University of Amsterdam, Netherlands; University of Arizona, USA; University of Bern, Switzerland; University of Cardiff, UK; University of Cologne, Germany; University of Ghent; University of Groningen, Netherlands; University of Leicester, UK; University of Leuven, Belgium; University of Stockholm, Sweden; Utah. The following National and International Funding Agencies
funded and supported the MIRI development:
NASA; ESA; Belgian Science Policy Office (BELSPO);
Centre Nationale d’Etudes Spatiales (CNES); Danish
National Space Centre; Deutsches Zentrum f\"ur Luft- und 
Raumfahrt (DLR); Enterprise Ireland; Ministerio
De Economalia y Competividad; Netherlands Research
School for Astronomy (NOVA); Netherlands Organisation
for Scientific Research (NWO); Science and Technology
Facilities Council; Swiss Space Office; Swedish
National Space Agency; and UK Space Agency.
E. D acknowledges support from the innovation and research Horizon 2020 program in the context of the  Marie Sklodowska-Curie subvention 945298 as well as from the Paris observatory-PSL fellowship.
M. G. is F.R.S.-FNRS Research Director. His contribution to this work was done in the framework of the PORTAL project funded by the Federal Public Planning Service Science Policy  (BELSPO) within its  BRAIN-be: Belgian Research Action through Interdisciplinary Networks program. 
P.-O.L.\, C.C., A.D.\, R.G.\, A.C.\ acknowledge funding support from CNES. 
T.G. and T.J.B.\ acknowledge support from NASA in WBS 411672.07.04.01.02.
O.A.\, I.A.\, B.V.\ and P.R.\ thank the European Space Agency (ESA) and the Belgian Federal Science Policy Office (BELSPO) for their support in the framework of the PRODEX Programme.
D.B.\ is supported by Spanish MCIN/AEI/10.13039/501100011033 grant PID2019-107061GB-C61 and and No. MDM-2017-0737. 
L.D.\ acknowledges funding from the KU Leuven Interdisciplinary Grant (IDN/19/028), the European Union H2020-MSCA-ITN-2019 under Grant no. 860470 (CHAMELEON) and the FWO research grant G086217N. 
I.K.\ acknowledges support from grant TOP-1 614.001.751 from the Dutch Research Council (NWO). 
O.K.\ acknowledges support from the Federal Ministry of Economy (BMWi) through the German Space Agency (DLR).
J.P.P.\ acknowledges financial support from the UK Science and Technology Facilities Council, and the UK Space Agency.
G.O.\ acknowledge support from the Swedish National Space Board and the Knut and Alice Wallenberg Foundation. 
P.T.\ acknowledges support by the European Research Council under Grant Agreement ATMO 757858. 
I.P.W.\ acknowledges funding from the European Research Council (ERC) under the European Union’s Horizon 2020 research and innovation programme (grant agreement No 758892, ExoAI), from the Science and Technology Funding Council grants ST/S002634/1 and ST/T001836/1 and from the UK Space Agency grant ST/W00254X/1.
F.A.M.\ has received funding from the European Union’s Horizon 2020 research and innovation programme under the Marie Sk\l{}odowska-Curie grant agreement no.\ 860470. 
E.D.\ has received funding from the European Union’s Horizon 2020 research and innovation programme under the Marie Skłodowska-Curie actions Grant Agreement no 945298-ParisRegionFP as well as from the Paris Observatory-PSL fellowship. 
G.V.L.\ acknowledges that some results of this work were partially achieved at the Vienna Scientific Cluster (VSC).
L.H.\ has received funding from the European Union’s Horizon 2020 research and innovation program under the Marie Sk\l{}odowska-Curie grant agreement no. 860470.
T.K.\ acknowledges funding from the KU Leuven Interdisciplinary Grant (IDN/19/028). 
L.C.\ acknowledges support by grant PIB2021-127718NB-100 from the Spanish Ministry of Science and Innovation/State Agency of Research MCIN/AEI/10.13039/501100011033. 
E.vD.\ acknowledges support from A-ERC grant 101019751 MOLDISK. 
T.P.R.\ acknowledges support from the ERC 743029 EASY. 
G.O.\ acknowledges support from SNSA.
P.P.\ thanks the Swiss National Science Foundation (SNSF) for financial support under grant number 200020\_200399.
O.A.\ is a Senior Research Associate of the Fonds de la Recherche Scientifique - FNRS.

 \item[Author Contributions Statement] 
All authors played a significant role in one or more of the following: development of the original proposal, management of the project, definition of the target list and observation plan, analysis of the data, theoretical modelling and preparation of this paper. Some specific contributions are listed as follows. 
P.-O.L.\ is PI of the JWST MIRI GTO European consortium program dedicated to JWST observations of exoplanet atmospheres; R.W.\ is co-lead of the program. 
E.D. provided overall program leadership and management of the TRAPPIST-1 b working group.
P.-O.L., T.G., J.B., T.H. made the design of the observational program and the setting of the observing parameters. 
A.D. generated simulated data for prelaunch testing of the data reduction methods. 
E.D., M.G (Gillon)., T.J.B. and P.-O.L.\ reduced the data, modeled the light curves and produced the eclipse depth. 
M.M., P.T.\ generated theoretical models to interpret the data. 
E.D. lead the writing of the paper, and M.G., T.J.B., P.-O.L., T.G., M.M., and P.M.\ made significant contributions to the writing of this paper. 
G.W.\ is the European PI of the JWST MIRI instrument,
P.-O.L., T.H., M.G (Guedel), B.V., L.C., E.vD., T.R., and G.Olofsson.\ are European co-PIs, and
O.A., A.M.G., I.K., L.D., R.W., O.K., J.P., G.Ostlin., D.R.\ and D.B.\ are European co-Is of the JWST MIRI instrument. T.G. is a U.S. co-I of the JWST MIRI instrument. A.G.\ led the MIRI instrument testing and commissioning effort.

\item [Competing interest statement]
The authors declare no competing interests
\end{addendum}

\newpage
\section*{Figures Legends/Captions (for main text figures)}

\textbf{Fig. 1 $\vert$ Phase folded JWST/MIRI observations of TRAPPIST-1 b.} \textbf{a.} Phase-folded light curve of the secondary eclipse of TRAPPIST-1 b at 12.8 $\mu m$, derived from the observations of 5 eclipses as part of GTO 1279 observations. \textbf{b.} Phase-folded light curve of the secondary eclipse of TRAPPIST-1 b at 15 $\mu m$, derived from re-analyses of the observations of 5 eclipses as part of GTO 1177. These figures are derived from the {\it ``fiducial analysis"} described in the Methods. Blue dots corresponds the corrected flux at the original time sampling, black dots shows the 10min binned corrected flux with an error equal to the standard deviation of the points within the bin, and the red curve show the eclipse model. \\

\noindent \textbf{Fig. 2 $\vert$ TRAPPIST-1 b's emission spectrum compared to bare-surface models.} Measurements of the eclipse depth of TRAPPIST-1 b in the 12.8 $\mu m$ and 15 $\mu m$ bands resulting from 5 visits in each band with their 1$\sigma$ uncertainties from our ``fiducial" joint analysis (MCMC analysis detailed in the Methods section), compared to realistic emission models for bare surface models from ref.~\cite{ih_constraining_2023}. The measurements are centered on the effective wavelength, which is computed by weighting the throughput of the filter with the corrected SPHINX synthetic stellar spectrum (see Methods for details). The error bar in wavelength stands for the width of the filter in each band. 
Colored markers show the band-integrated depth value for each model. Red and brown filled area show the response of the F1280W and F1500W filters respectively.  

\noindent \textbf{Fig. 3 $\vert$ Best fit atmosphere model:  an hazy-CO$_2$-rich atmosphere. } Eclipse spectrum (\textbf{a.}) and temperature structure (\textbf{b.}) for the haze atmosphere models. Model calculations are shown for haze mass fractions ($f_{haze}$ = 4.5E-4, $f_{haze}$ =4.5E-5, $f_{haze}$ =4.5E-7) as well as for our best-fit solution (blue line) and its 1$\sigma$ uncertainty interval (blue contour). On panel \textbf{a.}, the MIRI observations, resulting from 5 visits in each bands, are shown by black squares and their 1$\sigma$ errors derived from our ``fiducial" joint analysis are shown by black crosses. The measurements are centered on the effective wavelength, which is computed by weighting the throughput of the filter with the corrected SPHINX synthetic stellar spectrum (see Methods for details). The error bar in wavelength stands for the width of the filter in each band.

\clearpage
\section*{Extended Data Tables}

\begin{table*}[!h]
\centering
\caption{\textbf{Summary of the observations in JWST program GTO 1279.}}
\label{tab:observations}
\begin{tabular}{l|ccccc}\hline
                & visit 1       & visit 2       & visit 3      & visit 4 & visit 5      \\\hline
date            & 21 Nov. 2022 & 06 July. 2023 & 07 July. 2023 & 15 July. 2023 & 23 July. 2023 \\
start time      & 17:52:02              &  09:18:31             & 21:34:45            &  11:08:44        &  00:09:00       \\
end time        & 21:45:05              &  13:11:34             & 01:27:49             &  15:01:48     &  04:02:03          \\
duration (h)& 3.21          & 3.88          & 3.88        & 3.88       & 3.88   \\
$N_{int}$            & 280           & 315          & 315          & 315     & 315      \\
$N_{groups}$/int     & 17            & 15            & 15           & 15     & 15        \\
\end{tabular}
\end{table*}

\begin{table*}[h!]
\footnotesize
\centering
\caption{\textbf{Details of the different data reduction methods.}}
\label{tab:reductions}
\begin{tabular}{|p{2cm}|p{2.5cm}|p{2.5cm}|p{2.5cm}|p{2.5cm}|p{2.5cm}|}\hline
\textbf{Stage} & \textbf{Step} &  \textbf{ED reduction} & \textbf{MG reduction} & \textbf{TB reduction} & \textbf{POL reduction} \\
\hline
\textbf{Stage 1} & -STScI stage 1 run? & yes &yes & yes &No \\
    & & & & & \\
    & -Ramp weighting  &  default  & uniform &  default  & No ramp fitting  \\
    & & & & & \\
\hline
\textbf{Stage 2} & - Skip Photom? & yes & yes & yes & yes \\
& & & & & \\
\hline
\textbf{Stage 3} & - How to get centroid parameters? & 2D Gaussian fit to target & 2D Gaussian fit to target & 2D Gaussian fit to target & 2D Gaussian fit to target \\
        & & & & & \\
        & -Aperture shape & circle & circle & cirle & circle \\
        & & & & & \\
        & -Aperture radius & 5 pixels radius around the centroid at 12.8$\mu m$ and 15$\mu m$ &
            [3.5, 3.5, 3.5, 3.5, 3.5] pixels at 12.8$\mu m$ and [3.5, 3.8, 3.8, 3.8, 3.8] at 15$\mu m$ pixels around the center & 5 pixels radius at 12.8$\mu m$ and 8 pixels radius at 15$\mu m$ & 4 pixels radius \\
        & & & & & \\
        & -Background aperture size & 12-30 pixels for each visit & 30 - 45 pixels for each visit & 14--34 pixels & 30-45 pixels \\
        & & & & & \\
        & -Background subtraction method & subtracted the median calculated within the annulus  from the whole frame &subtracted the median calculated within the annulus  from the whole frame, after discarding 3$\sigma$ outliers & Subtracted the median within the annulus after first removing 5$\sigma$ outliers along the time axis & No background subtraction\\
        & & & & & \\
        & -Outlier rejection/time series clipping & sigma clipping set to 4, no exposure removed & 20min-4$\sigma$ median filtering, 3.5\% of exposures removed on average & Removed 3$\sigma$ outliers with respect to a 10-integration wide boxcar filter & outliers at 5 sigmas, 6 to 9 percents of integration discarded \\
        & & & & & \\
\hline
\hline
\end{tabular}
\end{table*}

\begin{table*}[h!]
\footnotesize
\centering
\caption{\textbf{Details of the different data analysis methods.}}
\label{tab:analysis}
\begin{tabular}{|p{3cm}|p{3.5cm}|p{3.5cm}|p{3.5cm}|}\hline
\textbf{Step} &  \textbf{ED analyses} & \textbf{MG analyses} & \textbf{TB analyses}  \\
\hline
-Sampling method & MH-MCMC & MH-MCMC & Nested Sampling using \texttt{dynesty}\cite{speagle_dynesty_2020}  \\
& & &  \\
-Setting & 2 chains of 100000 steps & 2 chains of 100000 steps  &  $N_{\rm dim}(N_{\rm dim}{+}1)//2$ live points, bounds=`multi', sample=`auto', $\Delta\log{\hat{\mathcal{Z}}}\leq0.1$ \\
& & &  \\
-1 chain to find correction factors & yes & yes & no  \\
& & &  \\
-Free planetary parameters & $(R_p/R_\star)^2$, $F_p/F_\star$, $b$, $e$, $t_0$ (for individual fits), TTVs (for join fits) & $(R_p/R_\star)^2$, $F_p/F_\star$, $b$, $e$, TTVs  & $F_p/F_\star$, $t_0$, $P$ (for joint fits)  \\
    & & &  \\
-Free stellar parameters & $R_\star$, $M_\star$, $T_{eff}$, [Fe/H] & $R_\star$, $M_\star$, $T_{eff}$, [Fe/H] & None  \\
    & & &  \\
-Potential polynomial on systematics & ramp, $x$, $y$, $FWHM_x$, $FWHM_y$  & ramp, $x$, $y$, $FWHM_x$, $FWHM_y$ & ramp, $FWHM_x$, $FWHM_y$, GP  \\
    & & &  \\
\hline
 -RMS of the residuals at 12.8$\mu m$ (ppm) & [711,	827,	728,	733,	882]  & [672,	784,	739,	724,	758] & [975,	883,	788,	912,	921] \\
& & &  \\
 -RMS of the residuals at 15$\mu m$ (ppm) & [1015,	892,	1028,	1250,	1041] & [965,	824,	859,	989,	899] & [1284,	1145,	1038,	1406,	1041] \\
& & &  \\
\hline
\end{tabular}
\end{table*}

\begin{table*}[h!]
        \footnotesize
        \centering
        \caption{\textbf{Derived eclipse depth with various approaches and fitting method.} \\ This Table provides all results from the various approaches and fits (individual, joint per band, joint with all eclipses).  }
        \label{tab:results_depths}
        \begin{tabular}{|p{3cm}|p{3cm}|p{3cm}|p{3cm}|p{3cm}|}\hline
            \textbf{Fit} &  \textbf{ED results} & \textbf{MG results} & \textbf{TB results} & \textbf{POL results} \\
            \hline
            \hline
            Individual fit for each 5 visits at 12.8$\mu m$ &  [$430 \pm 163$, $465 \pm 140$, $760 \pm 149$, $396 \pm 145$, $333 \pm 150$] & 
            [$433 \pm 173$, $482 \pm 198$, $793 \pm 139$, $305 \pm 134$, $332 \pm 169$]
            & [$460 \pm 340$, $600 \pm 180$, $690 \pm 220$, $530 \pm 150$, $310 \pm 220$] &  
            [$426 \pm 129$, $539 \pm 145$, $640 \pm 152$, $365 \pm 158$, $211 \pm 155$] \\
             & & & & \\
            \hline
            Individual fit for each 5 visits at 15$\mu m$ &  [$994 \pm 187$, $691 \pm 169$, $798 \pm 282$, $821 \pm 229$, $736 \pm 259$] & 
            [$771 \pm 176$, $632 \pm 252$, $822 \pm 207$, $815 \pm 200$, $524 \pm 189$]  & 
            [$790 \pm 210$, $510 \pm 210$, $950 \pm 170$, $820 \pm 220$, $829 \pm 200$]  &  
            [$806 \pm 177$, $585 \pm 174$, $780 \pm 181$, $758 \pm 176$, $776 \pm 164$]\\
             & & & & \\
            \hline
            \hline
            Joint fit of the 5 visits at 12.8$\mu m$ &  $463 \pm 87$  & $413 \pm 82$ &  $534 \pm 81$ & $436\pm70$  \\
             & & & & \\
            \hline
            Joint fit of the 5 visits at 15$\mu m$ & $823\pm 87$   & $718 \pm 91$ & $899 \pm99$ &  $741 \pm 80$\\
             & & & & \\
             \hline
             \hline
             Global joint fit of all 10 visits, 12.8$\mu m$ &  $431 \pm 90$  & $401 \pm 78$ & N/A & NA \\
             & & & & \\
            \hline
            Global joint fit of all 10 visits, 15$\mu m$ & $839 \pm 109$   & $718 \pm 88$  & N/A & NA \\
            & & & & \\
            \hline\hline
             ``fiducial" global joint fit at 12.8$\mu m$ &  \multicolumn{4}{c|}{{\centering 452 $\pm$ 86}} \\
            \hline
            ``fiducial" global joint fit at 15$\mu m$ &  \multicolumn{4}{c|}{{\centering 775 $\pm$ 90} } \\
             \hline
             \hline
        \end{tabular}
\end{table*}

\clearpage
\section*{Extended Data Figures}

\begin{figure}[ht!]
    \centering
    \includegraphics[width=0.8\textwidth]{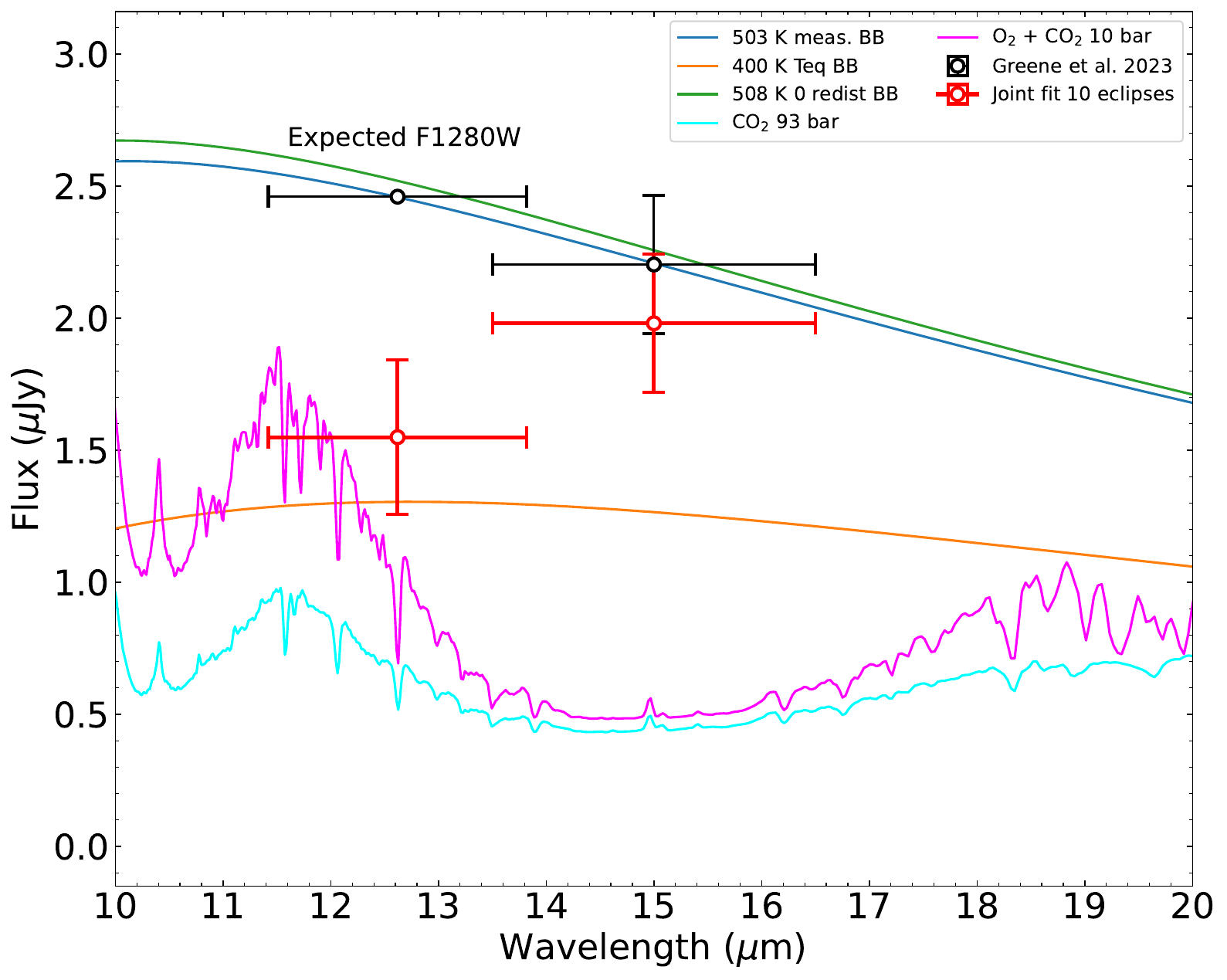}
    \caption{ \textbf{Planetary flux of TRAPPIST-1 b versus wavelength, from measurements and models.} The black dot with its error bar is the measured flux by ref.~\cite{greene_thermal_2023} from the analysis of 
    5 visits at 15$\mu m$, the black dot without an error bar is the expected flux in the F1280W filter for $T_{\rm B} = 503$ K. The planetary flux computed from this work is shown with red dots at 12.8 $\mu m$ and 15 $\mu m$. Fluxes are derived from our ``fiducial" joint analysis and the error bar in flux show their 1$\sigma$ uncertainties. The error bar in wavelength stands for the width of the filter in each band. The blue curve shows the blackbody curve expected for a dayside temperature of 503 K, the green curve shows the blackbody curve expected for 508 K apparent dayside temperature predicted for zero heat redistribution and no internal heating, and the orange curve shows the blackbody curve expected for $T_{\rm eq} = 400$ K temperature for isotropic redistribution of stellar heating. }
    \label{fig:planetary_flux}
\end{figure}

\begin{figure*}[ht!]
    \centering
    \includegraphics[width=0.5\textwidth]{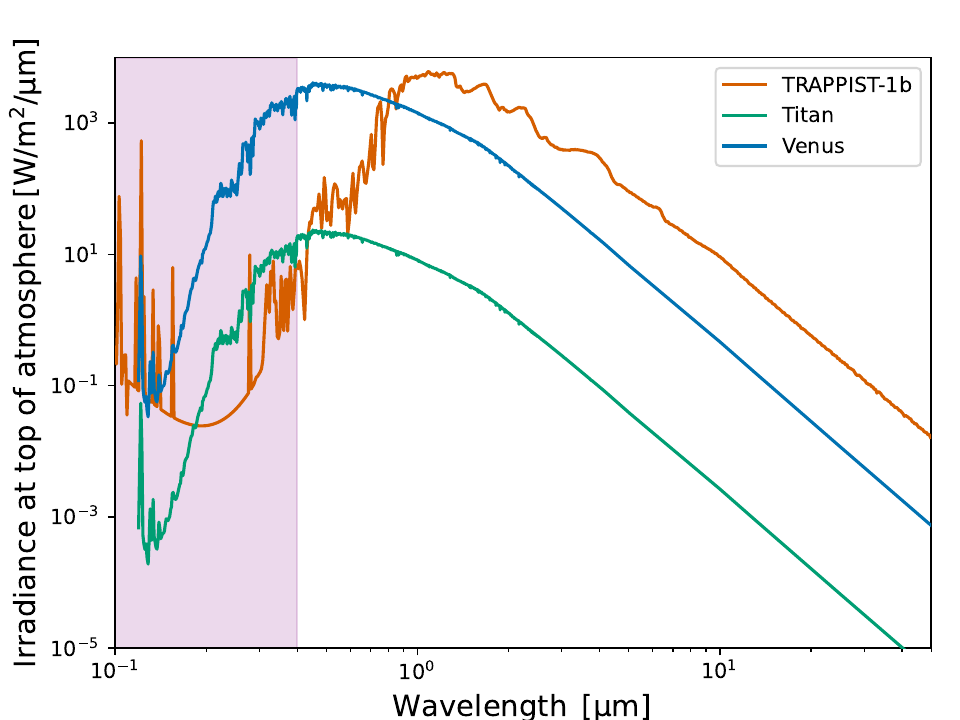}
    \caption{\textbf{Comparison of the irradiance at the top of the atmosphere for TRAPPIST-1b, Titan and Venus.} For the spectra of Titan and Venus we used the scaled standard Solar irradiance spectrum at air mass zero (AM0). For TRAPPIST-1b the spectrum from the Mega-MUSCLES project is used\cite{wilson_mega-muscles_2021}. The UV part of the spectrum is indicated by the violet band.}
    \label{fig:irradiance}
\end{figure*}

\end{methods}

\clearpage
\section*{References}

\newpage

\section*{Supplementary Figures}
\renewcommand{\figurename}{\hspace{-4pt}}
\renewcommand{\thefigure}{Supplementary Fig.~\arabic{figure}}
\renewcommand{\theHfigure}{Supplementary Fig.~\arabic{figure}}
\renewcommand{\tablename}{\hspace{-4pt}}
\renewcommand{\thetable}{Supplementary Table \arabic{table}}
\renewcommand{\theHtable}{Supplementary Table \arabic{table}}
\setcounter{figure}{0}
\setcounter{table}{0}

\begin{figure}[ht!]
    \centering
    \includegraphics[width=0.9\textwidth]{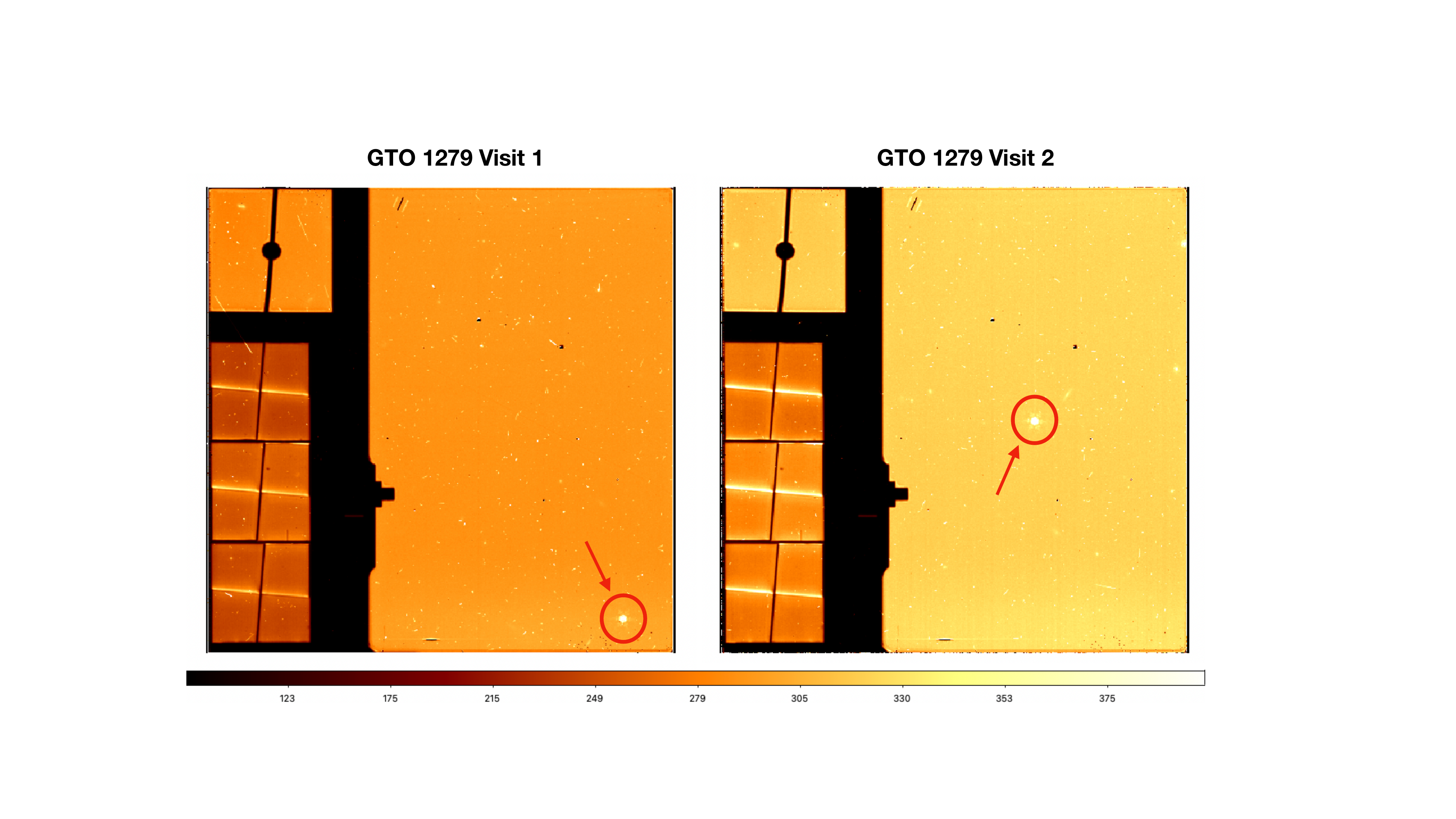}
    \caption{\textbf{STScI pipeline-corrected (Stage 2) image of the MIRI detector equipped with the F1280W filter for two distinct visits of program GTO 1279.} (\textbf{a.}) Detector image obtained with MIRI Imaging mode using the F1280W filter during the first visit of program 1279 when TRAPPIST-1 (in red) was offset. \textbf{b.} Image obtained with MIRI Imaging mode using the F1280W during the second visit when TRAPPIST-1 was centered in the field. Visits 3, 4 and 5 were carried out with the same setup as visit 2. MIRI imager focal plane is composed of several elements. The MIRI imager field of view (that we care for in program 1279) is the large array located on the right on the image for panel a. and b.. The left side of the imager was not used in our analysis, the top left sub-array corresponds to the Lyot coronoghraph, the three sub-arrays below are the three 4-quadrant phase masks coronographs. The color bar show the relative pixel intensity. }
    \label{fig:strategy}
\end{figure}

\newpage
\begin{figure*}[ht!]
    \centering
    \includegraphics[width=0.48\textwidth]{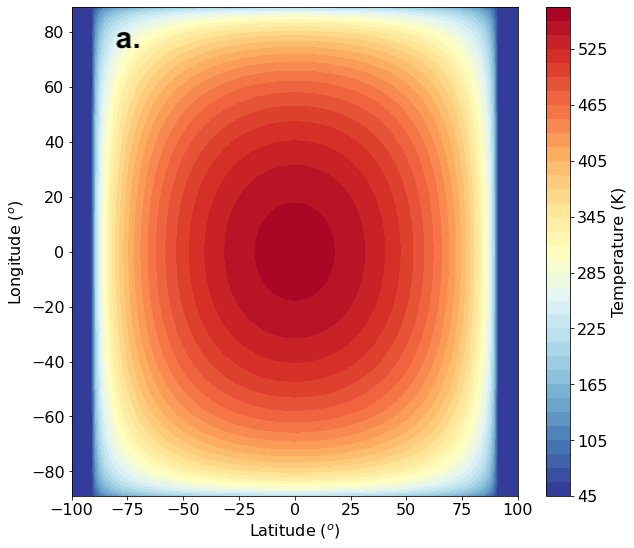}
    \includegraphics[width=0.47\textwidth]{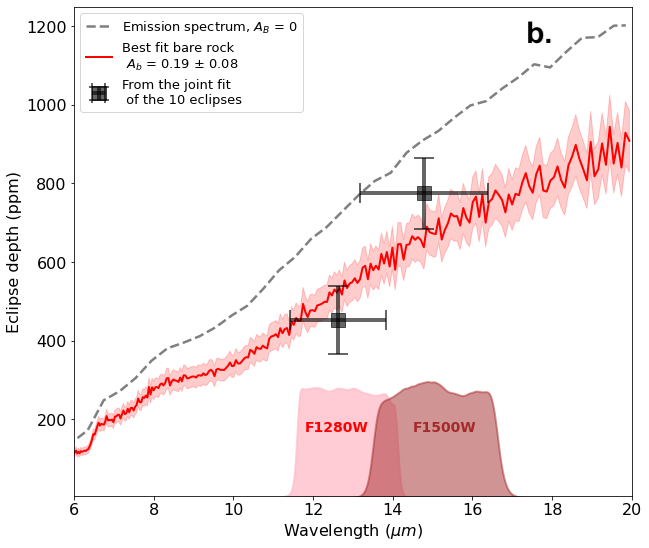}
    \caption{\textbf{Modeling of a theoretical black-body planet with variable albedo and comparison of its emission to JWST/MIRI measurements.} \textbf{a.} Temperature map of our blackbody emission model for an airless planet with no redistribution of heat. \textbf{b.} Best-fit albedo emission curve model derived from the fit of the two measurements and their 1$\sigma$ error for an airless planet modeled as a sum of blackbodies (red curve). The MIRI observations, resulting from 5 visits in each bands, are shown by black squares and their 1$\sigma$ errors derived from our ``fiducial" joint analysis are shown by black crosses. The measurements are centered on the effective wavelength, which is computed by weighting the throughput of the filter with the corrected SPHINX synthetic stellar spectrum (see Methods for details). The error bar in wavelength stands for the width of the filter F1280W and F1500W.}
    \label{fig:besfit_albedo}
\end{figure*}

\newpage
\begin{table*}[h!]
    \centering
    \begin{tabular}{|c|c|c|c|c|c|c|}
    \hline
         Surface model & Basaltic & Ultramafic & Fe-Oxidized & Granitoid & Grey \\
         \hline
         $\chi^2_r$ & 3.12 & 1.37 & 4.51 & 5.82 & 9.07\\
         \hline
    \end{tabular}
    \caption{$\chi_r^2$ statistics for the various bare surface models considering the two observed photometric measurements.  }
    \label{tab:chi2_surfaces}
\end{table*}

\newpage
\begin{figure*}[ht!]
    \centering
    \includegraphics[width=0.49\textwidth]{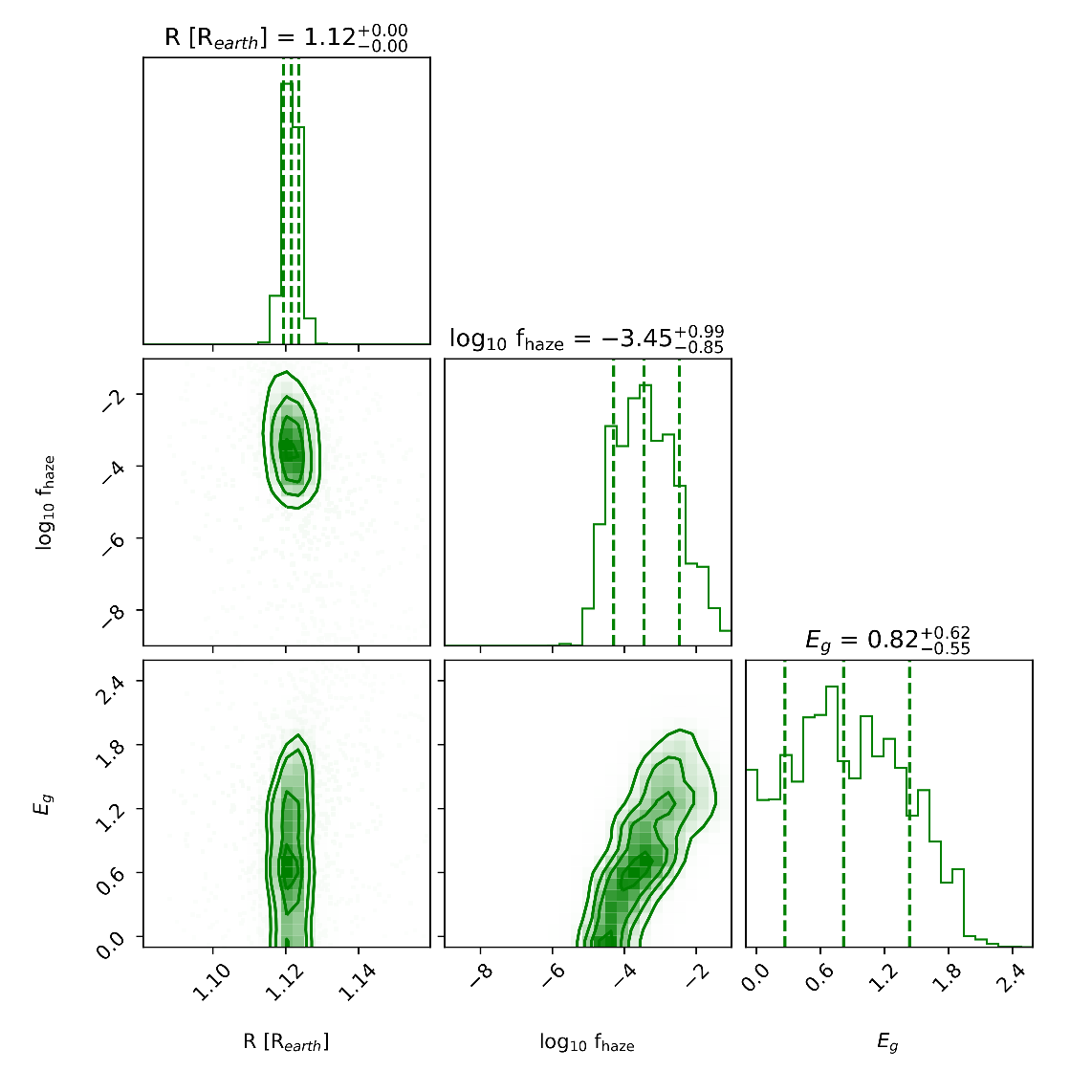}
    \includegraphics[width=0.5\textwidth]{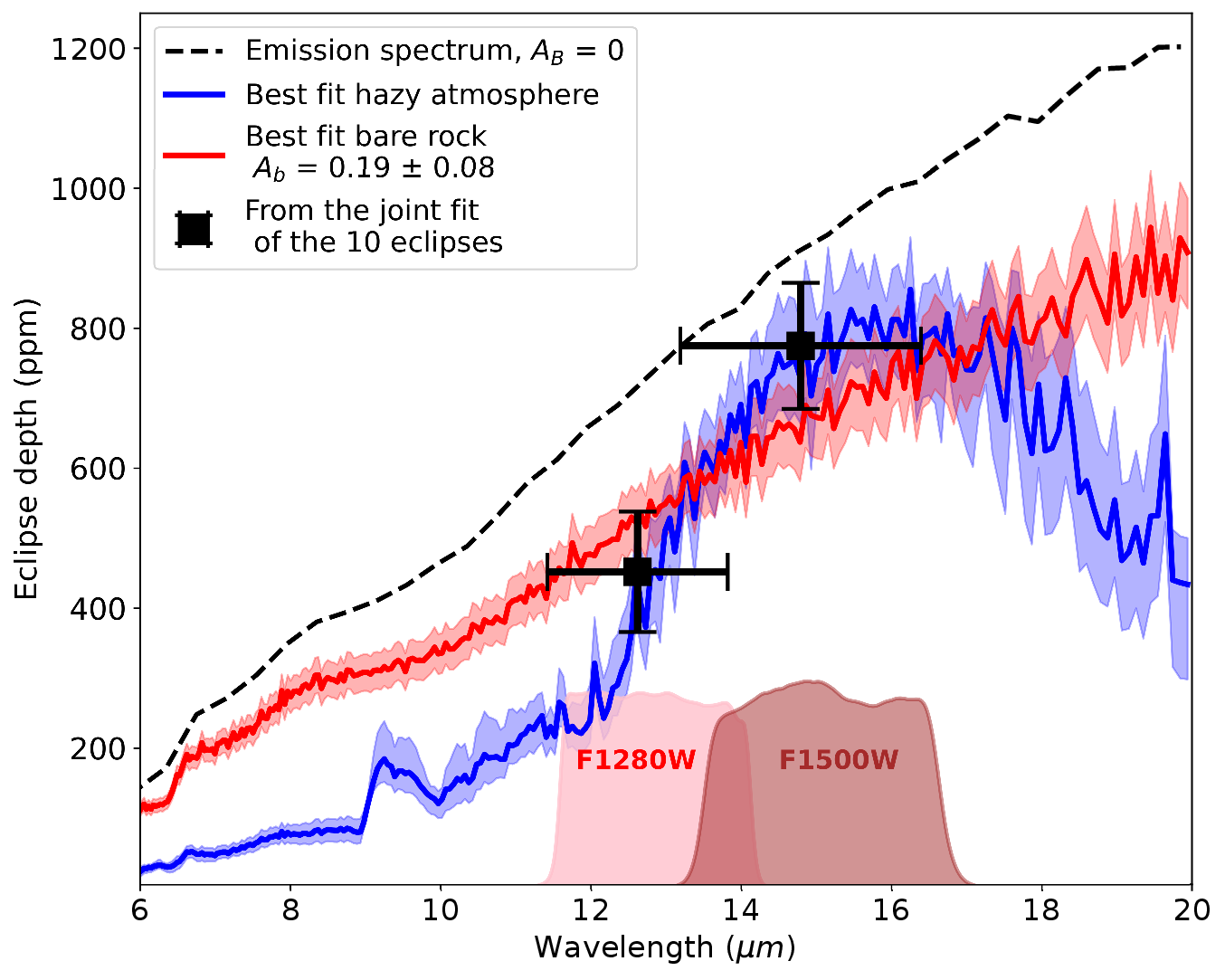}
    \caption{\textbf{Retrieval for an hazy-CO$_2$-dominated atmosphere.} \textbf{a.}  Resulting posterior distribution for the hazy atmosphere retrievals. \textbf{b.} Comparison of the measured eclipse depths with the best fit atmosphere model (blue  curve and 1-$\sigma$ error shade) and with the best fit bare-rock model (red curve and 1-$\sigma$ error shade). The eclipse depth derived from 5 visits in each bands are shown by black squares and their 1$\sigma$ errors derived from our ``fiducial" joint analysis are shown by black crosses.  The measurements are centered on the effective wavelength, which is computed by weighting the throughput of the filter with the corrected SPHINX synthetic stellar spectrum (see Methods for details). The error bar in wavelength stands for the width of the filter F1280W and F1500W.}
    \label{fig:retrieval_haze}
\end{figure*}

\newpage
\begin{figure*}[ht!]
    \centering
    \includegraphics[width=0.45\textwidth]{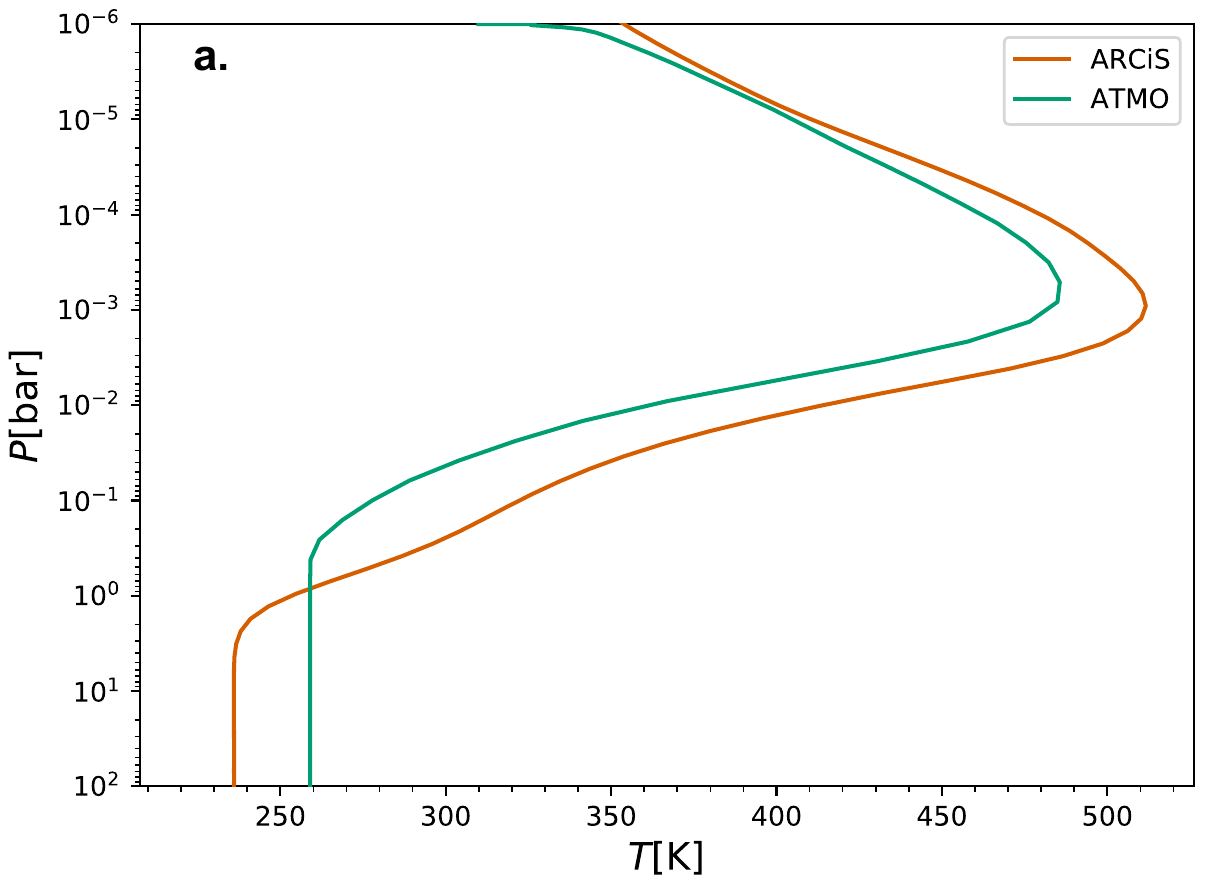}
    \includegraphics[width=0.45\textwidth]{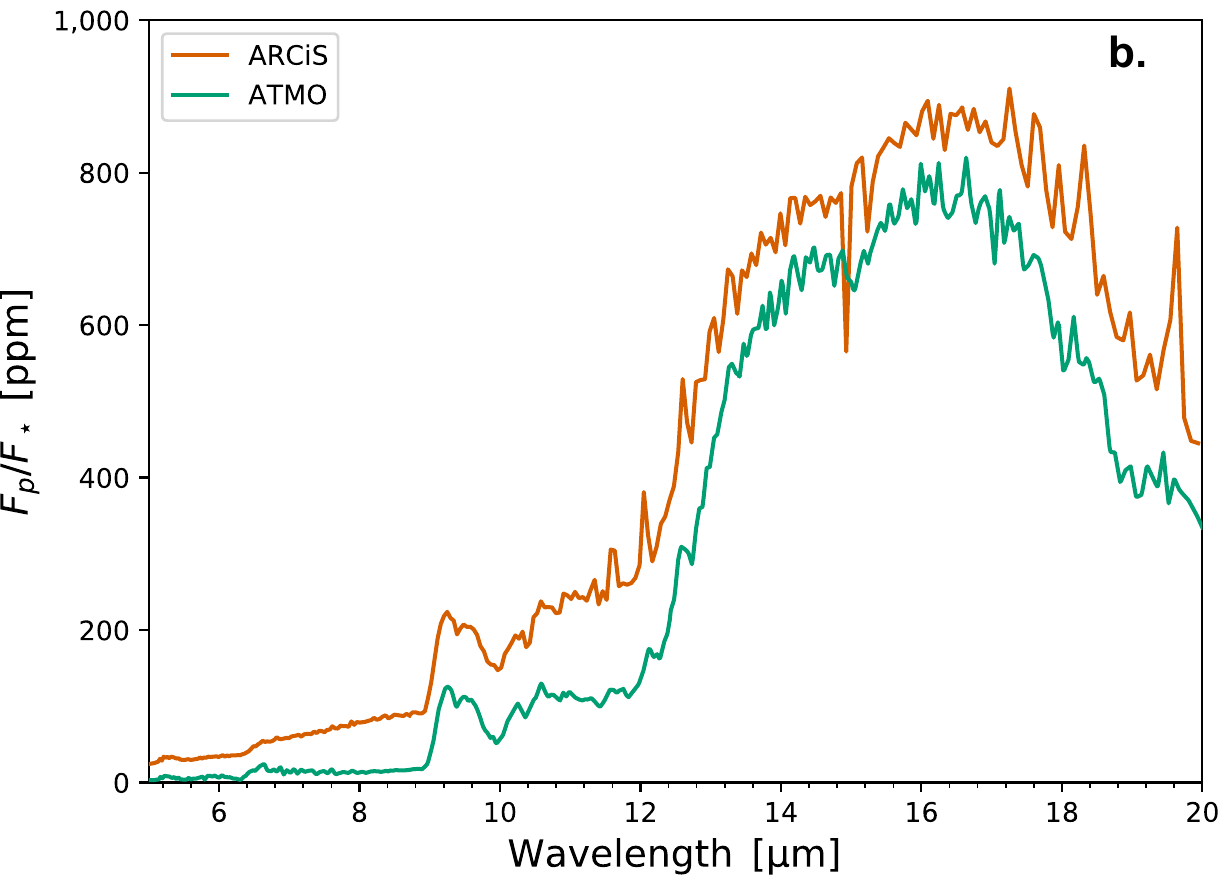}
    \caption{\textbf{Consistency of two distinct atmospheric modeling of the hazy-CO$_2$-dominated scenario.}  Comparison of the temperature structure (\textbf{a.}) and resulting eclipse depth (\textbf{b.}) for a similar atmospheric setup including CO$_2$ and parameterized haze opacity with the radiative transfer codes \texttt{ARCiS} and \texttt{ATMO}.}
    \label{fig:comparison}
\end{figure*}

\newpage
\begin{figure*}[ht!]
    \centering
    \includegraphics[width=0.99\textwidth]{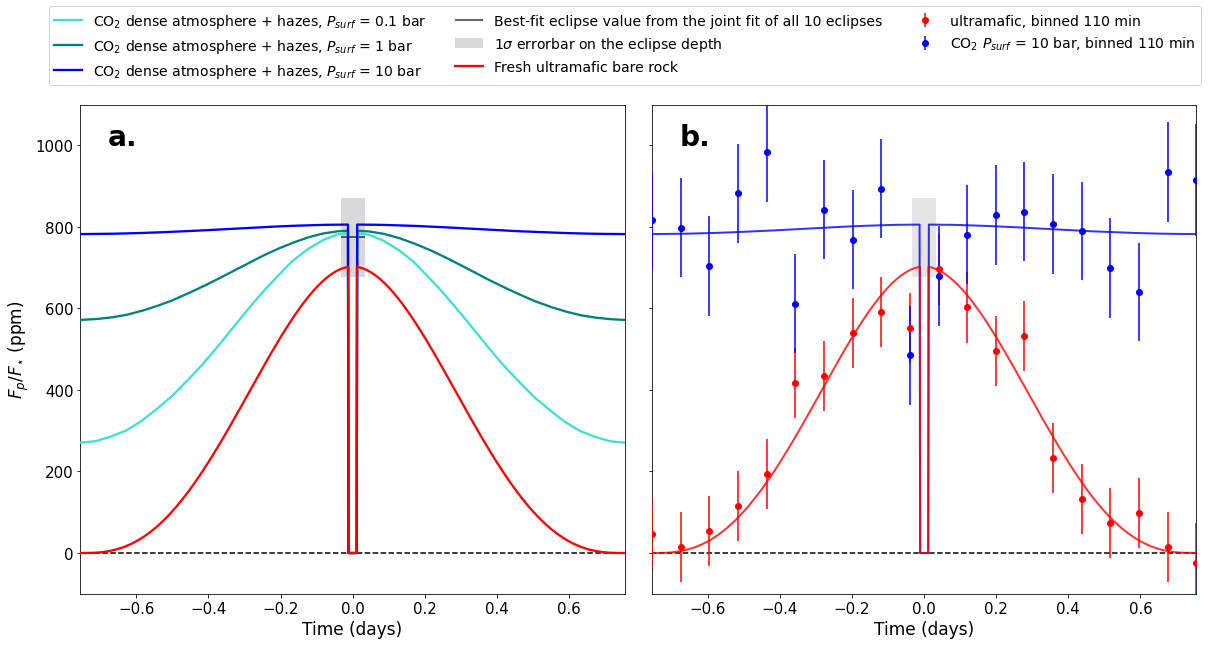}
    \caption{ \textbf{Phase curve simulations of TRAPPIST-1 b for a bare ultramafic rock scenario and hazy-CO$_2$-rich atmospheres.} \textbf{a.} Simulations of the phase curve of TRAPPIST-1 b at 15$\mu m$ for an airless planet with ultramafic surface and for a planet with a CO$_2$-rich atmosphere + hazes at various surface pressures, assuming full redistribution and no phase-curve offset. The gray patch shows for the eclipse depth measured at 15$\mu m$ (775 $\pm$ 90 ppm). \textbf{b.} Same but only for the two extreme cases and with simulations of MIRI F1500W data binned by 110 minutes over-plotted (red dots for ultramafic bare rock and blue dots for 10 bar hazy-CO$_2$-rich atmosphere),  with error bars equal to the standard deviation of the points within the bin. The observations of the secondary eclipses of TRAPPIST-1 b in the 12.8$\mu m$ and 15$\mu m$ bands did not allow to distinguish between these scenarios but the upcoming phase curve (GO 3077) should be able to.}
    \label{fig:phasecurves_haze}
\end{figure*}

\newpage
\begin{figure*}[ht!]
    \centering
    \includegraphics[width=1\textwidth]{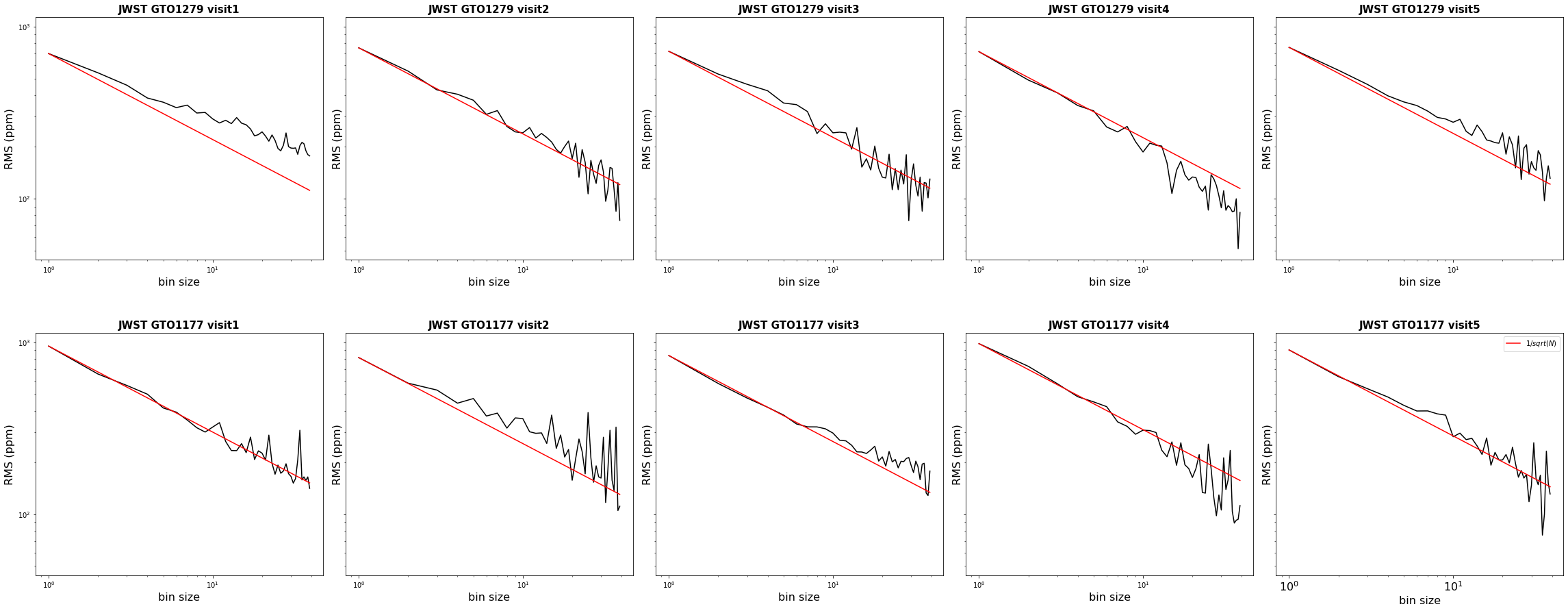}
      \caption{\textbf{Allan deviation plots.} Allan variance plots from the join fit of all 10 eclipses together. These plots show the evolution of the RMS of the best-fit residuals as a function of the bin size. The red line shows the decay slope in $1/\sqrt{N}$, where N is the bin size, which is expected in the absence of correlated noise. The fact that the RMS of our residuals follow the $1/\sqrt{N}$ slope proves that our fit of each visit (eclipse model + instrumental systematics) is satisfactory.  }
    \label{fig:allan_plots}
\end{figure*}

\newpage
\begin{figure*}[ht!]
    \centering
    \includegraphics[width=0.9\textwidth]{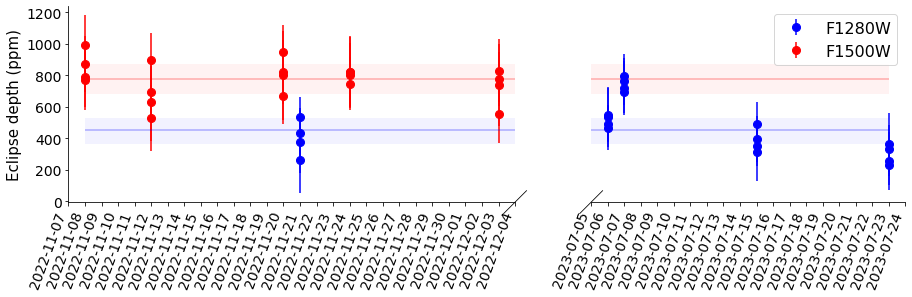}
    \caption{\textbf{Variability of the eclipse depth of TRAPPIST-1 b in time and wavelength.} Values of the secondary eclipse depth and their 1-$\sigma$ error bars derived from the four distinct data analysis approaches (MG, ED, TJB, POL) using individual fits of each visit at 12.8$\mu m$ and 15$\mu m$. The blue and red patches show the depth values derived from the global joint fit of all visits using the light curve model that lead to the smallest RMS of the residuals.  }
    \label{fig:depth_var_dates}
\end{figure*}


\newpage
\begin{figure*}[ht!]
    \centering
    \includegraphics[width=0.45\textwidth]{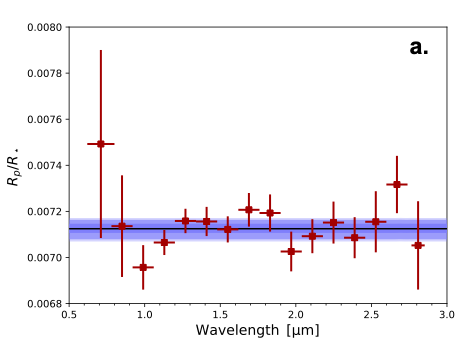}
    \includegraphics[width=0.45\textwidth]{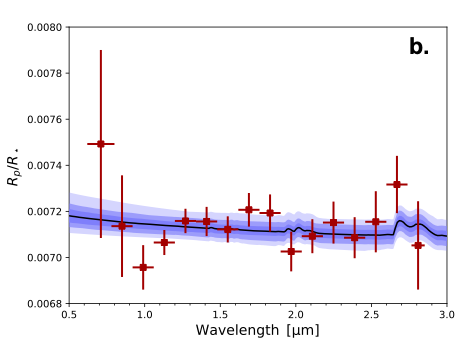}
    \caption{\textbf{Transmission spectrum of TRAPPIST-1 b.} The NIRISS observations from ref.~\cite{lim_atmospheric_2023} are shown in red dots, vertical error bars are the 1-$\sigma$ uncertainties, horizontal error bars represent the extent of each spectral bin. The blue curve shows the best fit for the bare rock scenario (\textbf{a.}) and the atmosphere scenario (\textbf{b.}).   }
    \label{fig:fit_transmission}
\end{figure*}

\begin{figure*}[ht!]
    \centering
    \includegraphics[width=0.81\textwidth]{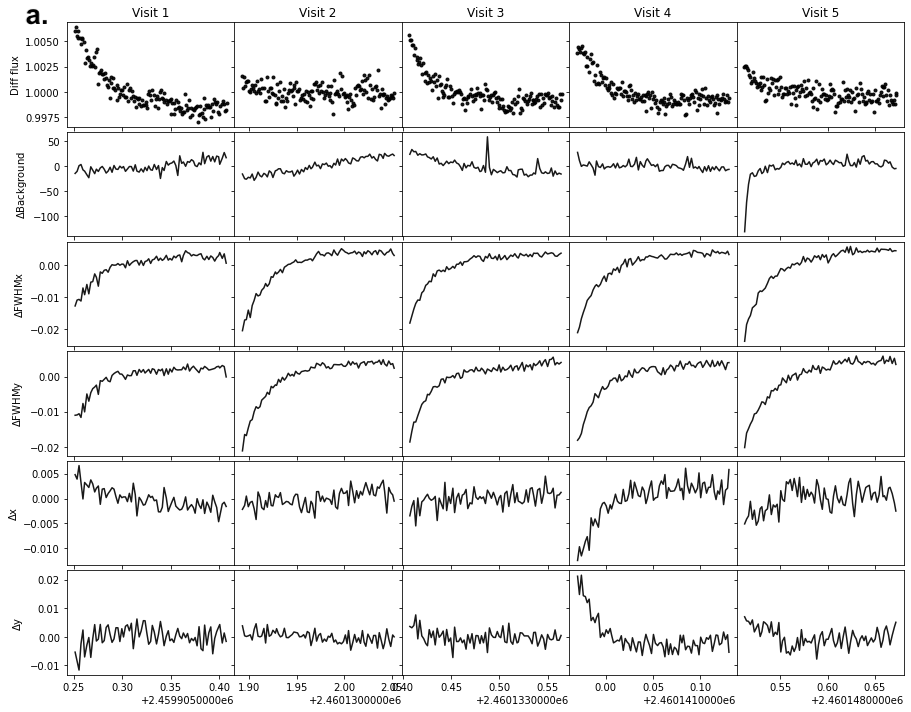}
    \includegraphics[width=0.81\textwidth]{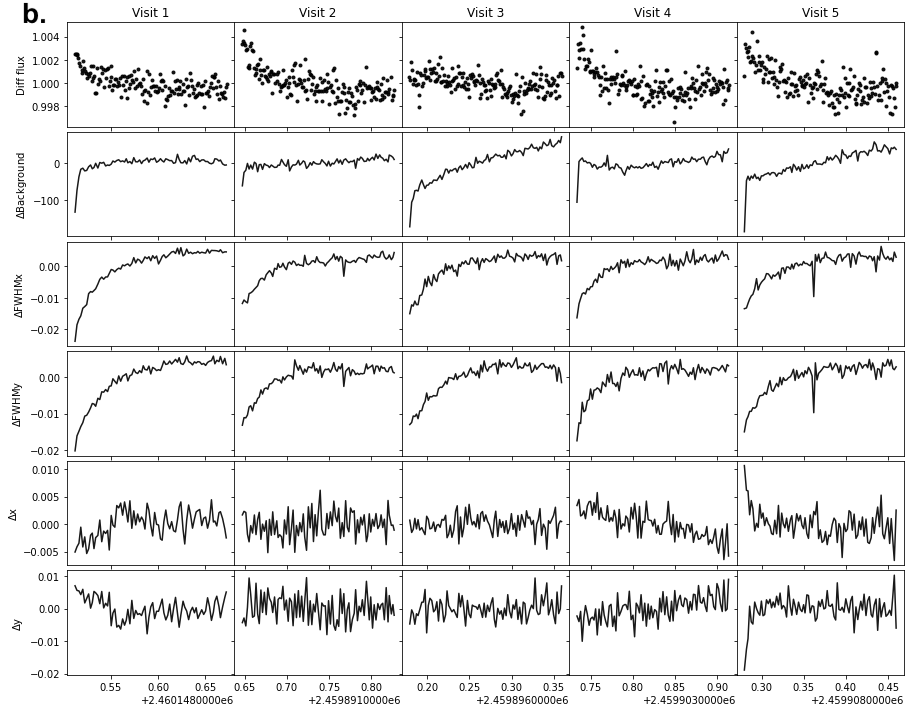}
    \caption{\textbf{Diagnostic plot of all 10 visits based on the MG reduction.} Evolution of the systematics over time for each visit of programs GTO 1279 (\textbf{a.}) and GTO 1177 (\textbf{b.}). }
    \label{fig:systematics_F1280W}
\end{figure*}


\begin{figure*}[ht!]
    \centering
    \includegraphics[width=1\textwidth]{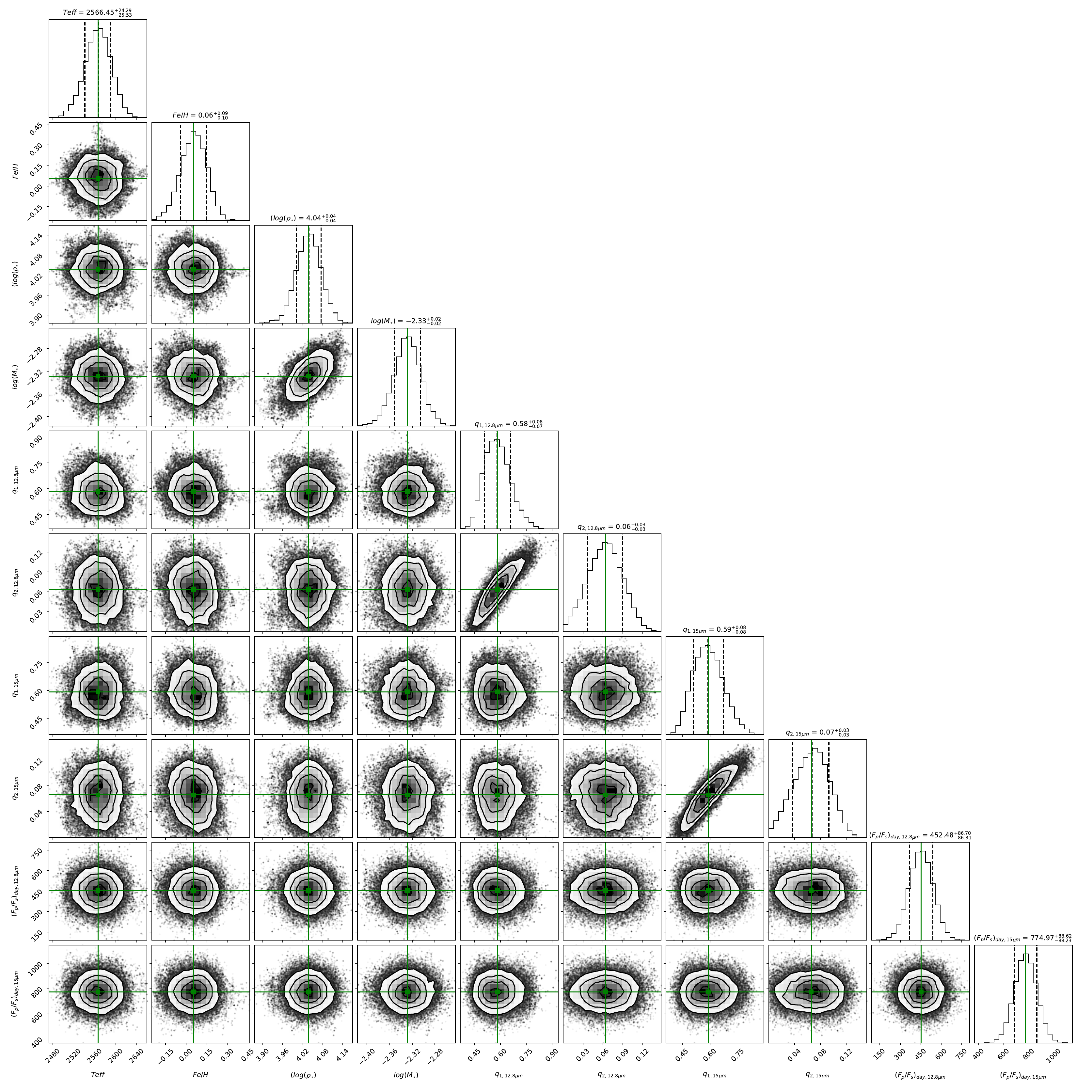}
    \caption{\textbf{Corner plot of the ``fiducial" fit.} We do not show the posteriors of the TTVs and eccentricities for clarity }
    \label{fig:corner}
\end{figure*}

\newpage

\end{document}